\documentclass[aps,prx,10pt,twocolumn,amsmath,amssymb,floatfix,superscriptaddress,prx]{revtex4-1}

\usepackage{graphicx}
\usepackage{txfonts}
\usepackage{braket}
\usepackage[colorlinks,citecolor=blue,linkcolor=blue,urlcolor=blue,plainpages=false,pdfpagelabels]{hyperref}

\newcommand{\pg}{g}
\newcommand{\hc}{\text{h.c.}}

\begin{document}

\title{Parafermion braiding in fractional quantum Hall edge states with finite chemical potential}

\author{Solofo Groenendijk}
\affiliation{Physics and Materials Science Research Unit, University of Luxembourg, L-1511 Luxembourg}
\author{Alessio Calzona}
\affiliation{Institute of Theoretical Physics and Astrophysics,
University of W\"urzburg, 97074 W\"urzburg, Germany}

\author{Hugo Tschirhart}
\author{Edvin G. Idrisov}
\author{Thomas~L.~Schmidt}

\affiliation{Physics and Materials Science Research Unit, University of Luxembourg, L-1511 Luxembourg}
\date{\today}

\begin{abstract}
Parafermions are non-Abelian anyons which generalize Majorana fermions and hold great promise for topological quantum computation. We study the braiding of $\mathbb{Z}_{2n}$ parafermions which have been predicted to emerge as localized zero-modes in fractional quantum Hall systems at filling factor $\nu = 1/n$ ($n$ odd). Using a combination of bosonization and refermionization, we calculate the energy splitting as a function of distance and chemical potential for a pair of parafermions separated by a gapped region. Braiding of parafermions in quantum Hall edge states can be implemented by repeated fusion and nucleation of parafermion pairs. We simulate the conventional braiding protocol of parafermions numerically, taking into account the finite separation and finite chemical potential. We show that a nonzero chemical potential poses challenges for the adiabaticity of the braiding process because it leads to accidental crossings in the spectrum. To remedy this, we propose an improved braiding protocol which avoids those degeneracies.
\end{abstract}
\maketitle

\section{Introduction}
Identical particles in two dimensions offer a more varied exchange statistics than merely the well-known bosonic and fermionic ones. The wave function of so-called anyonic particles can change by an arbitrary phase under the exchange of the particles. Moreover, if the particles span a degenerate ground state, they may even realize non-Abelian exchange statistics, where an exchange (``braiding'') implements a unitary transformation within the degenerate ground state subspace. The ground state can be used to encode quantum information, which can then be manipulated by braiding and can thus be used to implement logic gates for topologically protected quantum computation \cite{nayak08,Aasen16,alicea11,kitaev03}.

The most well-known excitation believed to offer non-Abelian braiding statistics is the Majorana bound state (MBS) which has received a lot of attention in the last decade \cite{kitaev01,wilczek09,heck12,beenakker13,franz13,alicea12,amorim15,sarma15,karzig15,karzig16,karzig19}. Strong experimental evidence for its existence has accumulated on several platforms \cite{willett13,nadjperge14,mourik12,albrecht16,deacon17}. MBSs are localized, but if kept at a finite distance, the exponential tails of their wavefunctions will overlap and the resulting hybridization lifts the exact ground state degeneracy \cite{sarma12,churchill13}. In superconducting nanowires, the characteristic dependence of the energy splitting on the chemical potential and the distance has become an important tool for identifying MBS \cite{sarma12,churchill13,maier14,zyuzin13,ChengMeng09},
\begin{equation}\label{eq:majorana}
\delta E \propto e^{-L/\xi}\cos(\mu L/ v_F).
\end{equation}
Here, $\xi$ is the superconducting coherence length, $L$ is the distance separating the two MBSs, $\mu$ is the chemical potential and $v_F$ the Fermi velocity. For the rest of this paper we will set $\hbar = 1$.

Despite offering non-Abelian braiding statistics, MBSs are unable to implement all logic gates necessary for universal quantum computation in a topologically protected way \cite{nayak08}. Therefore, non-Abelian anyons with a richer exchange statistics are needed. Recently, much attention has focused on realizing so-called $\mathbb{Z}_p$ parafermions in solid-state devices \cite{lindner12,clarke13,klinovaja15,alicea16,calzona18,chew18,mazza18,fleckenstein19,mazza2018,snizhko18tke,snizhko18cb,rossini2019,laubscher19}.
Although they do not allow for universal quantum computation, they permit to implement a richer set of quantum gates. Moreover, a network of such parafermions can be used as a building block for Fibonacci anyons, which would then indeed allow universal topological quantum computation \cite{mong14}.
An ordered set of parafermionic operators obeys the following operator algebra,
\begin{align}
\chi_k \chi_l &= e^{2 \pi i/p}\chi_l \chi_k,\qquad (\text{for }k>l) \notag \\ \chi_k^\dag &= \chi_k^{p-1},\quad \chi_k^p = 1.
\end{align}
In particular, MBSs can be regarded as $\mathbb{Z}_2$ parafermions. However, the richer exchange statistics for $p \geq 3$ makes $\mathbb{Z}_p$ parafermions more attractive for manipulating quantum information than MBSs~\cite{hutter16}.

Parafermions were proposed as localized zero-modes in fractional topological insulators (FTIs). The latter can be thought of as consisting of counterpropagating quantum Hall edge states with opposite spins and fractional filling factors $\nu = 1/n$ ($n$ odd) \cite{mengcheng12,lindner12,burnell16,clarke13,snizhko18tke,snizhko18cb,burrello13,milsted14}. Analogously to the situation in helical edge states of 2D topological insulators (2DTIs) \cite{fu09b}, a spectral gap in the pair of edge states can either be produced by allowing backscattering between the counterpropagating states or via the proximity effect of a superconductor (SC). These two mechanisms give rise to topologically different gaps, and the resulting interface zero-modes have $\mathbb{Z}_{2n}$ parafermions exchange statistics.

Further research has proposed their realization in different systems including 2DTIs \cite{orth15,vinkler17,zhang14,klinovaja14,fleckenstein19}, fractional topological SCs \cite{abolhassan13,laubscher19}, strongly interacting nanowires with spin-orbit coupling proximitized by a SC \cite{klinovaja14int,pedder17} or in bundles of strongly interacting nanowires \cite{klinovaja14bundle}. Recently, advances in experimental platforms based on the fractional quantum Hall effect, with a potential to host parafermions, have been achieved \cite{leonid18}.

\begin{figure}
\includegraphics[scale=0.5]{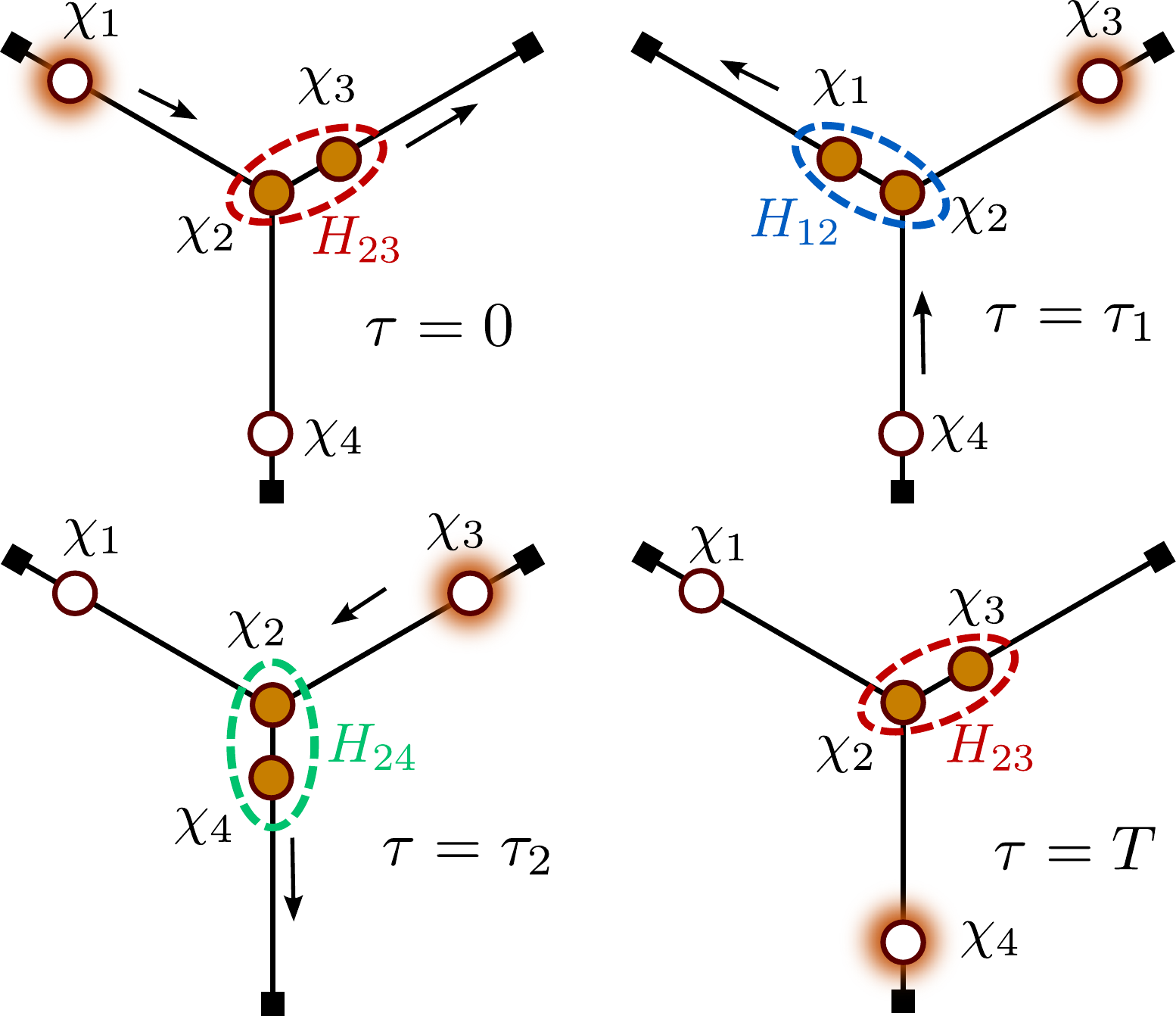}
\caption{Braiding protocol in 1D that exchanges parafermionic zero modes at times $0<\tau_1<\tau_2<T$. The black lines represent which parafermions are approaching each other, but do not represent actual interfaces. The empty circles represent zero-energy modes and the full circles represent finite energy modes.
Initially ($\tau = 0$), $\chi_{2,3}$ are nucleated at a finite energy, whereas $\chi_{1,4}$ are zero-energy states. At $\tau = \tau_1$, $\chi_{1,2}$ are then fused together such that the zero-energy state of $\chi_1$ gets transferred unto $\chi_3$. At $\tau = \tau_2$, we fuse $\chi_{2,4}$ such that the state of $\chi_4$ gets transferred unto $\chi_1$. Finally, at $\tau = T$, we fuse $\chi_{2,3}$ back into the vacuum and return to the initial configuration, albeit with the zero-energy states of $\chi_{1,4}$ exchanged.}
\label{fig:protocolA}
\end{figure}

When parafermions are generated in one-dimensional systems, it is difficult to braid them by adiabatically exchanging their position without bringing them together. In nanowire-based setups this problem might be circumvented by using T-junctions \cite{alicea11}, but for setups based on quantum Hall edge states, even this possibility does not exist. Proposals aimed at parafermion braiding are therefore based on repeated nucleation and fusion of parafermionic pairs \cite{lindner12,karzig15,bonderson08,bonderson13,knapp16}. One example of such a braid is shown in Fig.~\ref{fig:protocolA}, where the two parafermionic states $\chi_{1,4}$ are effectively braided using an auxiliary pair $\chi_{2,3}$ at an intermediate step. Mathematically, the braid shown in Fig.~\ref{fig:protocolA} can be modeled with a simple time-dependent Hamiltonian,
\begin{equation}\label{eq:braiding1}
H(\tau) = t_{23}(\tau)\chi_2\chi_3^\dag+t_{12}(\tau)\chi_1\chi_2^\dag+t_{24}(\tau)\chi_2\chi_4^\dag+\hc
\end{equation}
For $\mathbb{Z}_{n \geq 3}$ parafermions, the coupling coefficients $t_{ij}(\tau)$ can be complex and depend on the system parameters, like chemical potential, pairing gap, etc.  Braiding can then be understood as a closed loop in the parameter space of the Hamiltonian. Obviously, using such a braiding protocol makes it unavoidable to bring the parafermions within a finite distance of each other, and their interactions can then be understood as due to the overlap of their wavefunctions. Alternatively, parafermions can also be coupled via Coulomb interactions rather than wavefunction overlap \cite{snizhko18tke,snizhko18cb,burrello13,milsted14}.

Whenever parafermions are braided by fusion and nucleation, the notion of adiabaticity differs drastically from the case when braiding is performed by exchange of anyons in real space. In the latter case, two anyons can be physically exchanged while being kept far apart, e.g., two Majorana modes in a T-junction \cite{alicea11,sarma12}. Adiabaticity is then related to the superconducting gap of the system which separates the ground state manifold from the quasi-particle excitations.
For braiding based on nucleation and fusion, one needs to merge the parafermionic zero-modes with auxiliary parafermions, which inevitably induces finite-energy states below the superconducting gap.
Hence, in this case, braiding should be performed slowly compared to these smaller gaps, rather than the larger superconducting gap of the system.

In addition to these constraints, one should note also that in both cases, the zero-modes to be braided are not exactly at zero energy because their finite separation slightly lifts the ground state degeneracy. One needs to perform the braiding \emph{fast} compared to this exponentially small energy splitting, so that the ground state still appears degenerate. In principle, this sets an upper bound for the braiding time, but the latter becomes large for long systems.

Nevertheless, such a braid is still topologically protected in a weaker sense, namely in that the outcome is insensitive to small deformations of the path performed in the space of control parameters. According to Ref.~\cite{lindner12}, this path invariance holds if specific complex phases of the tunnel couplings $t_{ij}$,  which would give rise to accidental degeneracies, are avoided. However, Ref.~\cite{burnell16} has shown using a semi-classical approach that these phases are difficult to avoid when two parafermions are separated by a SC with a finite chemical potential or a backscattering region with a magnetic field. The reasons is that both cases give rise to a distance-dependent complex phase in the coupling amplitudes and to an energy splitting similar to Eq.~(\ref{eq:majorana}). In this work, we will assess the impact of these complex phases on braiding.

The article is organized as follows. In Sec.~\ref{sec:Summary}, we present a short overview of our main results. In Sec.~\ref{sec:FQH}, we use models for parafermions in fractional quantum Hall edge states or topological insulator edge states to derive the energy splitting as a function of distance and chemical potential. In Sec.~\ref{sec:numerical}, we simulate and compare two braiding protocols, both involving fusion and nucleation of parafermionic pairs, and compare their overlap in the presence of a nonzero chemical potential. In Sec.~\ref{sec:conclusion}, we present our conclusions.

\section{Summary and main results}\label{sec:Summary}

The main subject of our study are parafermions realized in a pair of quantum Hall edge states with opposite $g$-factors at filling factor $\nu = 1/n$ ($n$ odd). Such a system hosts $\mathbb{Z}_{2n}$ parafermions at the interfaces between regions gapped by either a SC or by backscattering \cite{lindner12,clarke13}. We first revisit the problem of calculating the overlap between two parafermions in such a system if they are separated by a finite distance. Using a combination of bosonization and refermionization, we find the expected exponential suppression of the ground state splitting with the distance, but we also show that an oscillating term occurs if the chemical potential is nonzero. This is in accordance with a recent result based on a semiclassical analysis \cite{burnell16}. Moreover, we show that no such oscillating term is present in the energy splitting when two parafermions are coupled via a backscattering region with a finite chemical potential.

Similarly, we investigate $\mathbb{Z}_4$ parafermions emerging as zero-modes in the helical edge states of a two-dimensional topological insulator. We show that our approach captures these systems as well, and find results which are analogous to those derived for quantum Hall edge states. This suggests that the form of the energy splitting (\ref{eq:majorana}) is a generic property of parafermions kept at a finite distance.

From this result, we deduce the effective coupling amplitudes between spatially separated parafermions, including the complex phases, as a function of their distance. These serve as the input parameters of the parafermionic toy model~(\ref{eq:braiding1}), which allows us to numerically simulate the braiding protocol proposed by Ref.~\cite{lindner12,clarke13}. Since the $\mathbb{Z}_4$ parafermion model is the simplest example capturing all the relevant physical phenomena, we choose this as a basis. However, we have checked that the results we obtained are valid as well for $\mathbb{Z}_{2n}$ ($n$ odd) parafermions.

For a system modelled by Eq.~(\ref{eq:braiding1}) with $\mathbb{Z}_4$ parafermions, having at least one vanishing coupling coefficient $t_{ij}$ at any given time ensures that the ground state is always at least $4$-fold degenerate. When at least one of the other $t_{ij}$'s is non-zero at a given time, a finite gap separates the ground-state manifold from the excited states. However, the width of this gap varies as a function of time throughout the entire braiding process, and the latter can be considered as adiabatic only if it is slow compared to the smallest gap reached during the entire process. In that case, the time-dependent change of parameters induces no transitions to states outside the ground state manifold. Importantly, one should note that the gap which determines whether or not the braid is adiabatic is related to the overlap between parafermions, and is thus much smaller than the superconducting gap, which separates the parafermionic states from the quasiparticle states. A more general case of braiding of non-Abelian anyons with pairwise interactions is discussed in Ref.~\cite{burrello13b}.

In the conventional braiding protocol, pairs of parafermions are repeatedly nucleated and braided in such a way that at a given time at most two parafermions are maximally coupled to each other. For such a protocol, we show that as the chemical potential increases to values $\mu L \gtrapprox 1$, where $L$ is the length of the system, the gap between the ground states and excited states tends to become exponentially small at certain points in the braiding process. In general, the adiabatic limit can therefore be reached only for very large braiding times.

We compare this with a modified braiding protocol, which is also based on nucleation and fusion, but in which up to three parafermions can be brought together and being maximally coupled. We show that this protocol offers larger gaps throughout the entire braiding process, so the adiabatic limit can be reached more easily.

\section{Coupling of parafermion modes in fractionalized systems}
\label{sec:FQH}

\begin{figure}
\includegraphics[width=\columnwidth]{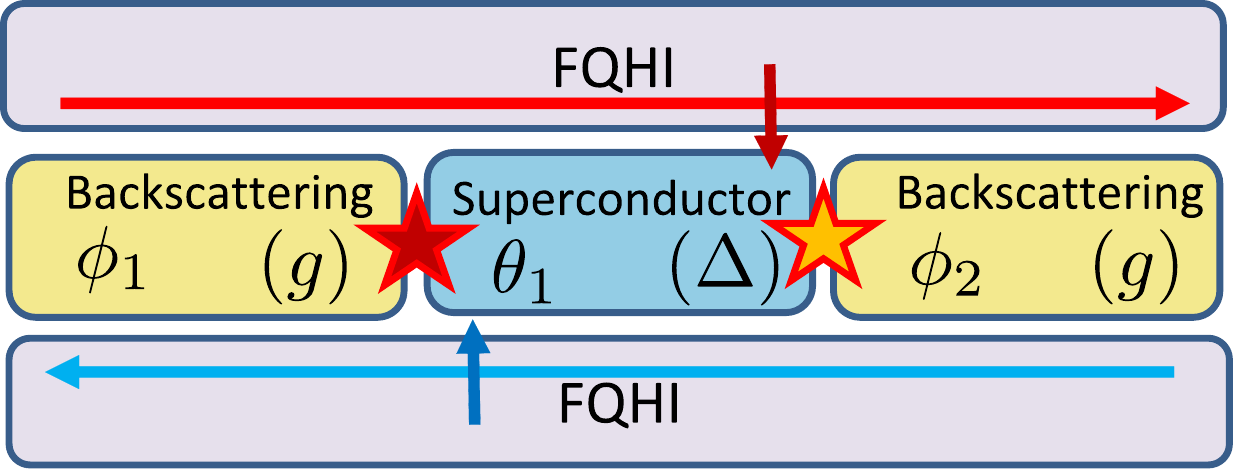}
\caption{Two FQH edge states with opposite spins, gapped out by either superconducting pairing or backscattering, can host $\mathbb{Z}_{2n}$ parafermions at the interfaces. If backscattering ($\pg$) dominates, the fields $\phi_{1,2}$ are pinned inside the backscattering region, whereas for dominant superconducting pairing ($\Delta$), the field $\theta_1$ is pinned.}
\label{fig:BSSCBSjunction}
\end{figure}

Let us consider two fractional quantum Hall (FQH) edge states with opposite $g$-factors, such that the system contains two coupled counterpropagating fractional edge states with opposite spins. A gap can then be opened either by allowing backscattering between the two edge states or by inducing Cooper pairing via the superconducting proximity effect (Fig.~\ref{fig:BSSCBSjunction}). The bosonized theory describing the effective 1D physics is given by the following double sine-Gordon Hamiltonian \cite{lindner12,clarke13},
\begin{align}\label{eq:lindner}
H =& \frac{v_F n}{2\pi}\int dx\Big\{[\partial_x\theta(x)]^2+[\partial_x\phi(x)]^2\Big\}-\frac{\mu}{\pi}\int dx\partial_x\phi(x) \nonumber \\
& -\int dx \frac{\Delta(x)}{n\pi a} \cos(2n\theta(x)) -\int dx  \frac{\pg(x)}{n\pi a} \cos(2n\phi(x)),
\end{align}
where $v_F$ is the Fermi velocity, $\Delta(x)$ the superconducting pairing potential which we take to be zero outside the superconducting region and constant $\Delta(x)=\Delta$ inside, and $\pg(x)$ is the backscattering strength which we take to be zero outside the backscattering region and constant $\pg(x)=\pg$ inside. The chemical potential $\mu$ is assumed to be constant throughout the system, the odd integer $n = 1/\nu$ is the inverse filling factor, and $a$ is a short-distance cut-off that regularizes the bosonization procedure. The bosonic fields $\theta(x)$ and $\phi(x)$ are related to the physical right-moving ($+$) and left-moving ($-$) electrons via the bosonization identity
\begin{equation}
\psi_{\pm}(x) = \frac{e^{-in\left[\pm \phi(x)-\theta(x)\right]}}{\sqrt{2\pi n a}},
\end{equation}
and fulfill the following fractional commutation relation
\begin{equation}
[\theta(x),\phi(y)]=\frac{i\pi}{n}\Theta(y-x),
\end{equation}
where $\Theta$ is the Heaviside function. As $n$ is odd, the physical electrons satisfy the correct anti-commutation relations. The chemical potential couples to the total charge density $\rho_C = \sum_{\alpha=\pm} \psi^\dag_{\alpha}\psi_\alpha  = \frac{1}{\pi}\partial_x\phi$. The spin density is given by $\rho_S = \sum_{\alpha=\pm}\alpha \psi^\dag_{\alpha}\psi_\alpha = \frac{1}{\pi}\partial_x\theta$. These operators are understood to be normal-ordered with respect to the ground state.

The system described by Eq.~(\ref{eq:lindner}) has a $2n$-fold degenerate ground state and hosts $\mathbb{Z}_{2n}$ parafermions at the interfaces between the backscattering and superconducting regions \cite{lindner12}. For strong pairing potential $\Delta$, the energy is minimized when the $\theta$ field is pinned to $\theta(x)=\pi\hat{k}_{\theta}/n$, where $\hat{k}_{\theta}$ is an integer-valued operator. Analogously, in the backscattering region the energy is minimized when the $\phi$ field is pinned to the value $\phi(x)=\pi\hat{k}_{\phi}/n$.

If neighboring parafermionic modes are kept at a finite distance, they can couple to each other either via the tunneling of quasiparticles across the superconducting region or the backscattering region. Since the braiding protocols we study involve bringing the parafermionic modes together and then separating them, our first aim is to investigate the effect of such tunneling.

The coupling between parafermions separated by a gapped region can be studied relatively easily if the coupling between a given pair of parafermions is much stronger than their coupling to all other parafermions. In the following, we will therefore study the coupling between (i) two parafermions separated by a superconducting region and (ii) two parafermions separated by a backscattering region.

\subsection{Tunneling across a superconductor}

We start by considering a single pair of parafermions which are separated by a short superconducting region. The Hamiltonian in the SC region reads
\begin{align}\label{eq:SCfqh}
H &=  \frac{v_F n}{2\pi}\int dx\Big\{[\partial_x\theta(x)]^2+[\partial_x\phi(x)]^2\Big\}-\frac{\mu}{\pi}\int dx\partial_x\phi(x) \nonumber \\
& -\frac{\Delta}{n\pi a} \int dx \cos(2n\theta(x)).
\end{align}
A system described by such a Hamiltonian is depicted in Fig.~\ref{fig:BSSCBSjunction}. This fractional sine-Gordon Hamiltonian with a chemical potential can be mapped onto an integer sine-Gordon Hamiltonian without chemical potential by defining the scaling and shift transformation
\begin{align}
\tilde{\theta}(x) &= n\theta(x) \notag \\
\tilde{\phi}(x) &= \phi(x)-\frac{\mu}{n v_F}x
\end{align}
which ensures that $[\tilde{\theta}(x),\tilde{\phi}(y)]=i\pi\Theta(y-x)$. After defining the Luttinger parameter $K=1/n$, the Hamiltonian reads (up to a constant term)
\begin{align}\label{eq:SCfqh2}
H  &=  \frac{v_F}{2\pi}\int dx\Big\{K[\partial_x\tilde{\theta}(x)]^2+\frac{1}{K}[\partial_x\tilde{\phi}(x)]^2\Big\}\nonumber \\& -\frac{\tilde{\Delta}}{\pi a} \int dx \cos(2\tilde{\theta}(x)).
\end{align}
where $\tilde{\Delta} = \Delta/n$. It is known that this sine-Gordon Hamiltonian exhibits a Berezinskii-Kosterlitz-Thouless (BKT) phase transition between a gapless Luttinger liquid phase and a gapped superconducting phase \cite{giamarchi}. Even tough repulsive interactions ($K < 1$) will suppress the superconducting term in the renormalization group (RG) sense, for large $\tilde{\Delta}$ the superconducting term will be dominant, and the system will be in the gapped phase. Electron-electron interactions described by the Luttinger parameter $K$ affect the gapped system only weakly, so we may take $K=1$ inside the quadratic term.

Within this approximation, the effective Hamiltonian~(\ref{eq:SCfqh2}) becomes diagonal in terms of new fermionic quasiparticles defined by $\tilde{\Psi}_{\pm} \propto \exp[-i(\pm \tilde{\phi}-\tilde{\theta})]$. We then rewrite the latter in terms of the slowly oscillating fields $\tilde{\psi}_\pm(x)$
\begin{equation}
\tilde{\Psi}_\pm(x) = e^{\mp i \tilde{\mu}x/v_F} \tilde{\psi}_{\pm}(x),\quad \tilde{\psi}_{\pm}(x) = \frac{1}{\sqrt{2\pi a}}e^{-i \left[\pm  \phi(x)- \tilde{\theta}(x)\right]},
\end{equation}
where $\tilde{\mu} = \mu/n$, and $[\tilde{\theta}(x),\phi(y)]=i\pi\Theta(y-x)$.
The procedure of first absorbing the chemical potential term in the bosonic fields before performing the approximation $K=1$ in the quadratic part of the Hamiltonian ensures that the bosonic theory can be refermionized, while keeping the correct scaling of the parameters by the Luttinger parameter $K=1/n$.

The effective Hamiltonian can now be rewritten as a fermionic Hamiltonian $H = H_{hl}+H_\mu + H_\Delta$, consisting of a helical liquid Hamiltonian $H_{hl}$, a chemical potential term $H_\mu$ generated by the oscillating term and a superconducting pairing Hamiltonian $H_\Delta$,
\begin{align}
H_{hl} &= -i v_F \sum_{\alpha=\pm} \alpha\int dx\ \tilde{\psi}_\alpha^\dag(x)  \partial_x \tilde{\psi}_\alpha (x), \\
H_{\mu} &= -\tilde{\mu} \sum_{\alpha=\pm} \int dx\ \tilde{\psi}_\alpha^\dag(x)\tilde{\psi}_\alpha (x), \\
H_{\Delta} &=  -\tilde{\Delta}\int dx\   \tilde{\psi}^\dag_+(x)\tilde{\psi}^\dag_-(x)  + \hc.
\end{align}
We can diagonalize this Hamiltonian by first going to momentum space, $\tilde{\psi}_{\pm}(x) = L^{-1/2} \sum_k e^{ikx}c_{k,\pm}$, and then defining the Bogoliubov transformation
\begin{align}
\begin{pmatrix}
c_{k,+}\\
c^\dag_{-k,-}
\end{pmatrix}
=
\frac{1}{\sqrt{2}} \begin{pmatrix}
\sqrt{1 + \epsilon_k/E(k)} & \sqrt{1 - \epsilon_k/E(k)} \\
-\sqrt{1 - \epsilon_k/E(k)} & \sqrt{1 + \epsilon_k/E(k)}
\end{pmatrix}
\begin{pmatrix}
\gamma_{k,+}\\
\gamma_{-k,-}^\dag
\end{pmatrix}.
\end{align}
The Bogoliubov quasiparticles obey the fermionic commutation relations $\{\gamma_{\sigma,k},\gamma^\dag_{\sigma',k'} \}=\delta_{k,k'}\delta_{\sigma,\sigma'}$ and give rise to the following diagonal Hamiltonian,
\begin{align}\label{eq:HSCeff}
H =& \sum_{k,\sigma}E(\sigma k)\gamma^\dag_{  k,\sigma}\gamma_{  k,\sigma},\\
E(k) =& \pm \sqrt{\varepsilon_k^2+\tilde{\Delta}^2},
\end{align}
where $\varepsilon_k= v_F k - \tilde{\mu} $. The information about the ground state energy splitting can be extracted from the zero-frequency limit of the retarded quasiparticle Green's function,
\begin{align}\label{eq:RGF}
\tilde{G}_{R}(x_2,x_1,t) =& -i\Theta(t)\left\langle\{\tilde{\psi}_{\pm}(x_2,t),\tilde{\psi}^\dag_{\pm}(x_1,0)\} \right\rangle\\
\tilde{G}_{R}(x_2,x_1,\omega) =& \int dt\ e^{i\omega t} \tilde{G}_{R}(x_2,x_1,t) ,\nonumber
\end{align}
where $\left\langle\cdots\right\rangle$ denotes the ground state expectation value with respect to the Hamiltonian~(\ref{eq:HSCeff}).
The time dependence of the quasiparticle operators are given by the Heisenberg equation of motion $\dot{\gamma}_{k,\sigma}(t) = i[H,\gamma_{k,\sigma}(t)]$ which is solved by
\begin{equation}
\gamma_{k,\sigma}(t) = e^{-iE(\sigma k)t}\gamma_{k,\sigma}(0).
\end{equation}
By expressing the operators in terms of Bogoliubov quasiparticles we have, after performing a shift of variables $k\to k- \tilde{\mu}/v_F$,
\begin{align}
\tilde{G}_{R}(x_2,x_1,\omega) =\nonumber \\ \frac{e^{i\tilde{\mu} L/v_F}}{ 2\pi  v_F}\int dk &\frac{(\omega + k)e^{i kL/v_F}}{\left(k+i\sqrt{\tilde{\Delta}^2-\omega^2}\right)\left(k-i\sqrt{\tilde{\Delta}^2-\omega^2}\right)},
\end{align}
where $L=(x_2-x_1)>0$ is the distance separating the parafermions. Since we are interested only in energies $\omega \ll \tilde{\Delta}$, the denominator is well behaved. Because $L>0$ we can integrate over momenta using the residue theorem with a contour in the upper half part of the complex plane.
The Green's function can then be written as
\begin{align}
  \tilde{G}_R(x_2,x_1,\omega) =
 -\frac{e^{-L/\xi_\Delta(\omega)}e^{i\tilde{\mu} L/v_F}}{ 2v_F}\Bigg[i+\frac{\omega \xi_\Delta(\omega)}{v_F}\Bigg],\nonumber
\end{align}
where we defined the correlation length $\xi_\Delta(\omega)=v_F/(\tilde{\Delta}^2-\omega^2)^{1/2}$. The zero-energy contribution to the nonlocal density of states is thus
\begin{equation}\label{eq:tunnelSC}
-\frac{1}{\pi}\text{Im} \tilde{G}_R(x_2,x_1,0) \propto \exp\left(-\frac{\Delta L}{n v_F}\right)\cos\left(\frac{\mu L}{nv_F}\right).
\end{equation}
The oscillating term due to the finite chemical potential $\mu$ is the main result of this analysis, and is in agreement with the results for the semi-classical limit in Ref.~\cite{burnell16}.

An investigation of a similar sine-Gordon Hamiltonian as Eq.~\eqref{eq:SCfqh2} using the exact form factors in order to compute the tunneling density of states suggests that the exponential suppression in Eq.~\eqref{eq:tunnelSC} is indeed independent of the Luttinger parameter in the quadratic part of Eq.~\eqref{eq:SCfqh2} and depends only on $v_F$ and the prefactors of the cosine \cite{pedder16b}.

Our results in Eq.~(\ref{eq:tunnelSC}) thus generalizes the well-known exponentially decaying and oscillatory overlap of Majorana bound states in superconducting nanowires. Indeed, Majorana fermions correspond to $n=1$, in which case we recover Eq.~(\ref{eq:majorana}) exactly.

\begin{figure}
\includegraphics[width=\columnwidth]{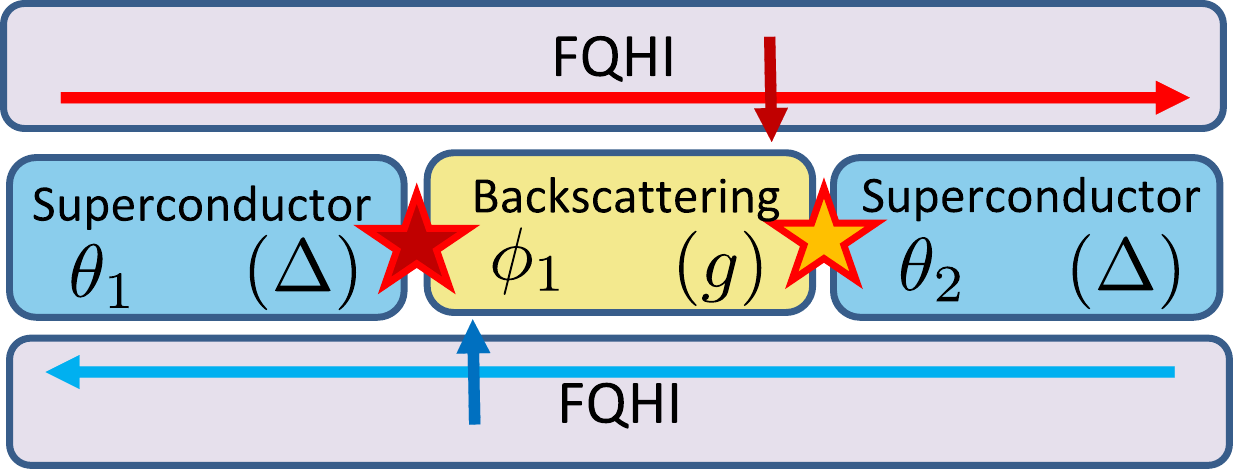}
\caption{Two FQH edges with opposite spin polarizations, gapped by either a SC or backscattering region, hosting $\mathbb{Z}_{2n}$ parafermion at the interfaces between the SC and the backscattering region. Such opposite spin polarization can be realised using 2DEG with opposite g-factors \cite{clarke13}. For strong SC and backscattering interaction, the fields $\theta_{1,2}$ and $\phi_1$ are pinned in their respective regions.}
\label{fig:SCBSSCjunction}
\end{figure}

\subsection{Tunneling across a backscattering region}

Next, we compare the previous results to the system depicted in Fig.~\ref{fig:SCBSSCjunction}, i.e., a pair of parafermions separated by a short backscattering region. Defining the scaling transformation $(\theta,n\phi)=(\tilde{\theta},\tilde{\phi})$ we can rewrite the Hamiltonian in the backscattering as a sine-Gordon Hamiltonian with Luttinger parameter $K=n$:
\begin{align}\label{eq:SCfqh3}
H & = \frac{v_F}{2\pi}\int dx\Big\{K[\partial_x\tilde{\theta}(x)]^2+\frac{1}{K}[\partial_x\tilde{\phi}(x)]^2\Big\}-\frac{\mu}{\pi K}\int dx\partial_x\tilde{\phi}(x) \nonumber \\
& -\frac{g}{K\pi a} \int dx \cos(2\tilde{\phi}(x)).
\end{align}
As before, we can approximately use $K=1$ in the quadratic terms since we assume the system is already gapped by the backscattering term. Hence, Eq.~(\ref{eq:SCfqh3}) is quadratic in terms of the fermionic quasiparticle operators $\tilde{\psi}_{\pm}(x)$ and can be diagonalized by a similar Bogoliubov transformation.

However, the role of the chemical potential in the Green's function turns out to be different because the chemical potential term now contains the same operator as the cosine term. Indeed, one finds for the retarded Green's function, with the renormalized parameters $\tilde{\pg}=\pg/K=\pg/n$ and $\tilde{\mu}=\mu/K=\mu/n$,
\begin{equation}
\tilde{G}_R(x_2,x_1,\omega)= \frac{1}{2\pi v_F}\int dk \frac{(\omega-\tilde{\mu}+k)e^{ikL/v_F}}{k^2+ \tilde{\pg}^2-(\omega-\tilde{\mu})^2-2i\omega\delta },
\end{equation}
where $L=x_2-x_1 > 0$. For $\omega,\tilde{\mu} \ll \tilde{\pg}$, one can use $\delta \to 0$ in the denominator. The fraction can be split into a sum of two fractions with one having two poles of order one at $\pm i[ \tilde{\pg}^2-(\omega-\tilde{\mu})^2]^{1/2}$ and the other one having a pole at $+ i [ \tilde{\pg}^2-(\omega-\tilde{\mu})^2]^{1/2}$. Since the integrals converge in the upper half part of the complex plane, we can use the residue theorem. The imaginary part of resulting retarded Green's function reads, using the physical parameters of the theory,
\begin{equation}
-\frac{1}{\pi}\text{Im} \tilde{G}^{(0)}_R(x_2,x_1,0) \propto
\exp\left(-\frac{\sqrt{\pg^2-\mu^2}}{v_F n}L\right).
\end{equation}
In contrast to Eq.~(\ref{eq:tunnelSC}), the chemical potential here affects only the effective gap seen by the parafermions and hence the length scale of the exponential decay. However, it does not give rise to an oscillating term. The form of the Green's function is in agreement with Ref.~\cite{pedder16b}.

To achieve an oscillatory term in the splitting of two parafermions separated by a backscattering region, one would have to add an effective Zeeman-like term $\propto B_z (\partial_x \theta)$ instead of the chemical potential term. This can be seen via the duality transformation $(\Delta,\theta,\phi,\mu)\leftrightarrow (\pg,\phi,\theta,B_z)$ of the Hamiltonian~(\ref{eq:SCfqh}).

\subsection{\texorpdfstring{$\mathbb{Z}_{4}$}{Z4} parafermions in topological insulator edge states}

Before concluding this section, we would like to apply the same approach to another important class of parafermions which do not fall into the category of $\mathbb{Z}_{2n}$ ($n$ odd) parafermions discussed so far.

As stated in the introduction, $\mathbb{Z}_4$ parafermionic models are among the simplest systems capturing the relevant physical phenomena. The same refermionization approach can also be used to determine the coupling strengths of $\mathbb{Z}_4$ parafermions appearing in helical edge states of two-dimensional topological insulator. Such time-reversal symmetric systems can be engineered in a physical setup similar to that in Figs.~\ref{fig:BSSCBSjunction} and \ref{fig:SCBSSCjunction}, albeit with electronic instead of fractionalized edge states. We thus also need to consider two-particle backscattering rather than one-particle backscattering:
In such systems, time-reversal symmetry forbids single-particle backscattering, but two-particle backscattering, which converts two right-movers into two-left movers, is allowed and opens a gap in the helical edge state spectrum. We can describe the 1D theory of interfaces between superconducting and backscattering regions in such an edge state by the following model Hamiltonian \cite{orth15}
\begin{align}
H &= \frac{v_F}{2\pi}\int dx \left\{K [\partial_x\theta(x)]^2+\frac{1}{K}[\partial_x\phi(x)]^2 \right\}\nonumber\\
&-\int dx \frac{\Delta(x)}{\pi a}\cos\left(2\theta(x)\right)  -\int dx  \frac{g(x)}{\pi a}\cos\left(4\phi(x)\right) \nonumber\\
&-\frac{\mu}{\pi}\int dx \partial_x\phi(x).
\end{align}
Here, $K$ is the Luttinger parameter, which quantifies the strength of the electron-electron interactions inside the helical edge state. Moreover, $g(x)$ is the effective two-particle backscattering strength, $\Delta(x)$ is the superconducting pairing amplitude, and $v_F$ is the Fermi velocity. As before, we will assume that each region of the edge state is dominated either by $g$ or $\Delta$. The bosonic fields $\phi$ and $\theta$ obey the usual commutation relation $[\theta(x),\phi(y)]=i\pi \Theta(y-x)$. The chemical potential $\mu$ is taken to be constant everywhere and couples to the total particle density $\rho_C = \frac{1}{\pi}\partial_x\phi $.

Let us first consider two parafermions separated by a superconducting region. The superconducting pairing term is diagonal in the physical fermions. Writing the Hamiltonian in this basis inside the superconducting region, one recovers a quadratic Hamiltonian. Deep in the gapped phase, the RG-marginal electron-electron interactions can be neglected, and one finds the following zero-energy nonlocal density of states,
\begin{equation}\label{eq:resultSCZ4}
-\frac{1}{\pi}\text{Im} G^{(0)}_R(x_2,x_1,0) \propto e^{-  \frac{\Delta L}{v_F}}
\cos\left(\frac{\mu L}{ v_F}\right).
\end{equation}
This result is in fact identical to the corresponding result for MBS in Eq.~(\ref{eq:majorana}).

In the opposite limit of two parafermions separated by a two-particle backscattering region, we start by defining the canonical transformation $(\tilde{\phi},\tilde{\theta})= (2\phi,\theta/2)$. As before we can then diagonalize the backscattering term by introducing fermionic quasiparticles defined by the boson fields $\tilde{\phi}$ and $\tilde{\theta}$. Deep in the gapped phase for strong backscattering strength, the theory becomes non interacting in terms of the new fermions at the Luther-Emry point $K=1/4$. The zero-energy nonlocal density of states becomes
\begin{equation}\label{eq:FM}
-\frac{1}{\pi}\text{Im}G^{(0)}_R(x_2,x_1,0) \propto e^{-\sqrt{g-(\mu/2)^2}L/v_F},
\end{equation}
where again no oscillations are present. This corresponding result for $\mu = 0$ was derived before using a different approach in Refs.~\cite{pedder16,pedder17}.

\subsection{Coupling parameter between lattice parafermions}
The goal of this section is to make the connection between
the coupling coefficients $t_{ij}$ in the lattice Hamiltonian~(\ref{eq:braiding1}) and the retarded nonlocal Green's function we calculated previously. This will allow us to make the coupling constant's dependence on the systems parameters (chemical potential, coupling strengths, length etc) explicit.

We consider the low-energy physics of parafermion modes separated by a superconductor in either a fractionalized system or a helical system. In the former case the localized modes are $\mathbb{Z}_{2n}$ parafermions, and in the latter $\mathbb{Z}_{4}$ parafermions.
The physics can be described by the following effective parafermion tight binding Hamiltonian \cite{alicea11,clarke13,lindner12,burnell16}
\begin{equation}
H = t\chi_1\chi_2^\dag + t^*\chi_2\chi_1^\dag,
\end{equation}
where $t=|t|e^{i\varphi}$ denotes the effective complex coupling coefficient. As we showed in previous sections, the two parafermions are coupled via the tunneling of quasiparticles across the junction. The eigenstates of this Hamiltonian can be labeled by the integers $q=0,\cdots,p-1$ using
\begin{equation}
\chi_1\chi_2^\dag\ket{q} = -e^{2\pi i\left(q+\frac{1}{2}\right)/p}\ket{q},\quad \chi_2\chi_1^\dag\ket{q} = -e^{-2\pi i\left(q+\frac{1}{2}\right)/p}\ket{q}.
\end{equation}
The spectrum of the Hamiltonian can then be written as
\begin{equation}\label{eq:spectraPF}
E(q) = -2|t|\cos\left[\frac{2\pi}{p}\left(q+\frac{1}{2}\right)+\varphi\right].
\end{equation}
Since the nonlocal retarded Green's function $\tilde{G}^R(x_2,x_1,\omega)$ in the limit $\omega\to 0$ contains the information about the probability amplitude for tunneling of quasiparticles between the points $x_2$ and $x_1$, as well as the phases picked up by the tunneling, we identify the Green's function with the hopping amplitudes of the parafermion lattice model, i.e.,
\begin{equation}\label{eq:toycoupling}
|t| \propto \begin{cases}
e^{- \Delta L/nv_F} & (\mathbb{Z}_{2n}) \\
e^{- \Delta L/v_F} &  (\mathbb{Z}_{4}) \\
\end{cases},\quad \varphi = \begin{cases}
\mu L/nv_F & (\mathbb{Z}_{2n} ) \\
\mu L/v_F &  (\mathbb{Z}_{4} )\\
\end{cases}.
\end{equation}

\section{Numerical simulation of braiding}\label{sec:numerical}

In this section we will discuss the effect of a finite chemical potential on the braiding of parafermions using the expressions for the hopping amplitudes. In particular, we will consider the braiding of two parafermions $\chi_1$ and $\chi_4$ performed by nucleating an additional parafermionic pair $\chi_2$ and $\chi_3$. The whole system is thus described by Hamiltonian \eqref{eq:braiding1},
\begin{equation}\label{eq:braiding1a}
H(\tau) = t_{23}(\tau)\chi_2\chi_3^\dag+t_{12}(\tau)\chi_1\chi_2^\dag+t_{24}(\tau)\chi_2\chi_4^\dag+\hc.
\end{equation}

To fully capture the effect of a chemical potential on braiding we will only consider coupling coefficients of the form Eq.~\eqref{eq:toycoupling}. A similar form can be obtained by replacing the superconducting by a backscattering region, and the chemical potential by a Zeeman-like term.
The braiding scheme we are going to investigate has been proposed and studied in several papers \cite{lindner12,clarke13,chew18}, but the detrimental effect of a finite chemical potential has so far not been assessed. Furthermore, we propose a way to mitigate this effect by a judicious choice of the time-dependence of the parameters $t_{ij}$.

During the braiding process, the coupling strengths are varied in time by controlling the distances $L_{ij}(\tau)\in[L_{min},L_{max}]$ between parafermions $\chi_i$ and $\chi_j$.
The minimal model for the coupling strengths with distance-dependent phases has the form
\begin{equation}
\label{eq:coupling}
t_{ij}(\tau) = e^{i\mu L_{ij}(\tau)}e^{-L_{ij}(\tau)}e^{i\varphi_0},
\end{equation}
where $\varphi_0$ is a global phase. In the following, we will measure all length scales in units of $n v_F / \Delta$ and energies are in units of $\Delta$.
We will focus on $\mathbb{Z}_4$ parafermions, as this is the simplest example which captures the relevant physics. We stress, however, that the results will be qualitatively similar for other $\mathbb{Z}_{2n}$ parafermions.

 \subsection{Braiding by nucleation and fusion}

Before presenting the numerical simulations, let us discuss the general features of braiding implemented by repeatedly fusing and nucleating pairs of parafermions in a 1D system \cite{clarke13,lindner12}. Such a process corresponds to an adiabatic change of the parameters of the Hamiltonian (\ref{eq:braiding1a}) along a closed loop in parameter space, and acts as a non-trivial unitary transformation within the degenerate ground-state manifold. Interestingly, if one considers only specific classes of loops which obey some particular constraints, the unitary transformations which can be obtained form a $4$-dimensional representation of the braid group \cite{orth15, karzig15}.

Let us be more specific by focusing on the braiding of parafermions $(\chi_{1},\chi_4)$. At the beginning, they are far apart from each other and therefore completely decoupled, while the nucleated parafermions $\chi_{2}$ and $\chi_3$ are strongly coupled. During the braiding process, we want to avoid situations in which the four parafermions are all close together and we thus require that, at each step, at least one parafermion must be far away from the others. The presence of such a decoupled parafermion ensures at least a four-fold degeneracy of the ground-state manifold, since there is at least one parafermionic operator which commutes with the Hamiltonian (\ref{eq:braiding1a}). In the case where all the parafermions are coupled, both adiabatic and non adiabatic errors may occur, as studied in the case of Majorana zero-modes \cite{heck12,Sekania17,fulga13}.

The Hilbert space of the whole system being $4\times 4=16$ dimensional, one can in principle encounter additional degeneracies, e.g., when all the couplings are switched off ($t_{12} = t_{23} = t_{24} = 0$), and the ground-state manifold becomes $16$ dimensional. The appearance of such additional degeneracies must also be avoided throughout the whole braiding process, as it would make it impossible to implement an adiabatic evolution of the system.

\begin{figure}
	\centering
	\includegraphics[scale=0.38]{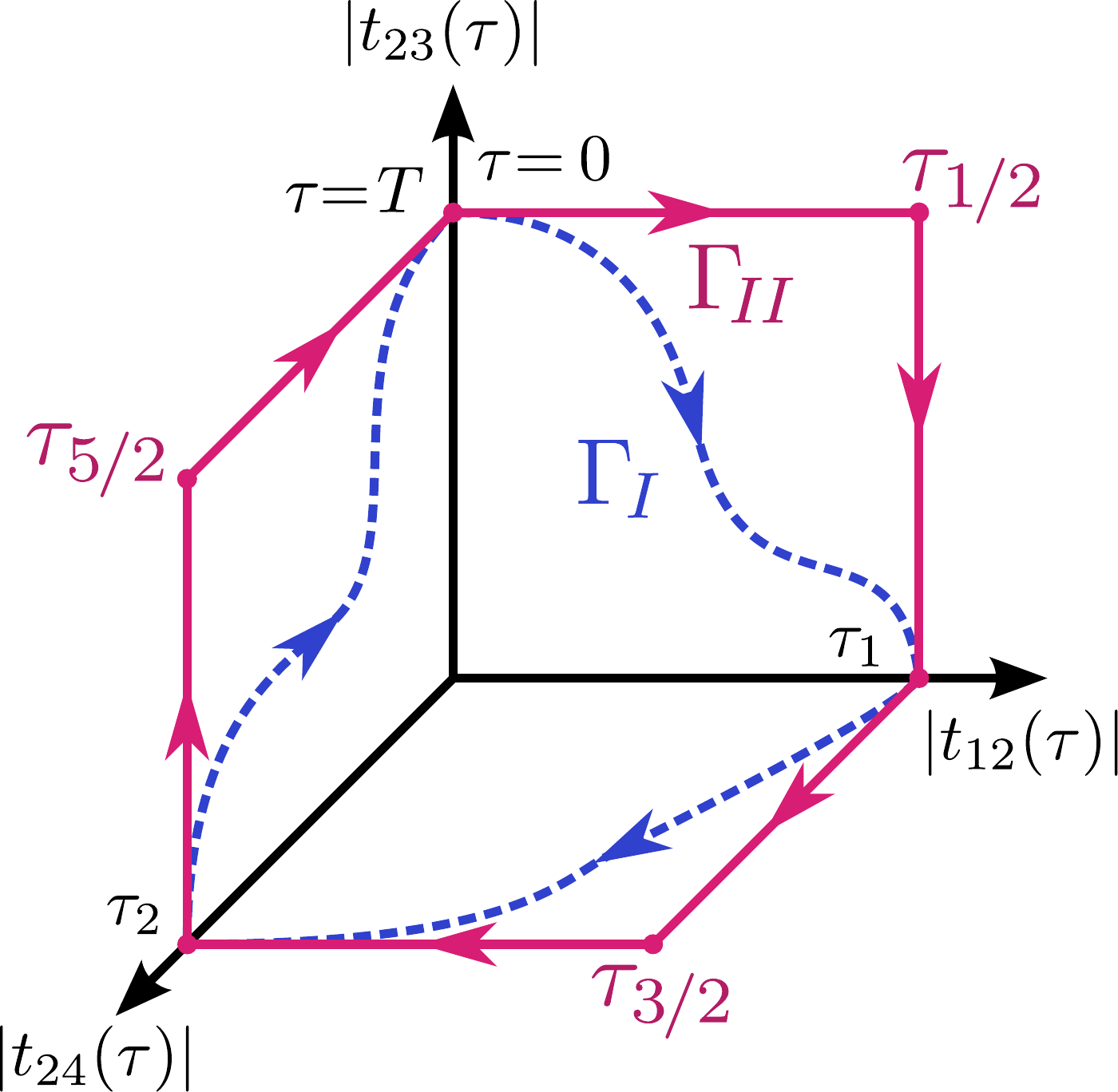}
	\caption{Two loops $\Gamma_I$ and $\Gamma_{II}$ in the configuration space spanned by the coupling constants $|t_{23}|$, $|t_{12}|$ and $|t_{24}|$. The loop $\Gamma_I$ passes through the points $|t_{23}|=1$, $|t_{12}|=1$, $|t_{24}|=1$ and $|t_{23}|=1$ at the times $\tau = 0< \tau_1 <\tau_2<T$ respectively. In the loop $\Gamma_{II}$ we add the intermediate stages where $|t_{23}|=|t_{12}|=1$, $|t_{12}|=|t_{12}|=1$ and $|t_{24}|=|t_{23}|=1$ at the times $\tau_{1/2}<\tau_{3/2}< \tau_{5/2}$.}
	\label{fig:braidingloops}
\end{figure}

Loops which satisfy these constraints, i.e., at least one decoupled parafermion and no additional degeneracies, are associated with unitary transformations on the ground state manifold which form a representation of the braid group. Two notable examples are provided in Fig.~\ref{fig:braidingloops}, where we plot the loops $\Gamma_{I}$ and $\Gamma_{II}$ in the $3$-dimensional parameter space spanned by $|t_{ij}|=e^{- L_{ij}}$. Pictorial representations of the main stages of loops $\Gamma_{I}$ and $\Gamma_{II}$ are given in Fig.~\ref{fig:protocolA} and \ref{fig:protocolB} respectively. By inspecting the parafermion position in these figures, it is clear that the two oriented loops  $\Gamma_{I}$ and $\Gamma_{II}$ both lead to a clockwise braid of $\chi_{1}$ and $\chi_4$. An anticlockwise braid can be achieved by implementing their time-reversed partners. As a clockwise braid cannot be continuously deformed into an anticlockwise one, a loop cannot be smoothly transformed in its time-reversed partner without violating the two constraints discussed above.
The braiding of parafermions by nucleation and fusion can still be considered topologically protected since its outcome is insensitive to small changes of the physical parameters $L_{ij}$, provided the four-fold degeneracy is maintained throughout the braiding process.

While different loops such as $\Gamma_{I}$ and $\Gamma_{II}$ are completely equivalent in terms of the braiding outcome, in the following we will show that they can differ a lot when it comes to their actual implementation. In particular, the requirements posed by adiabaticity may vary dramatically. The reason for that traces back to the presence of complex phases in the coupling amplitudes $t_{ij}$. As already pointed out in Ref.~\cite{lindner12}, particular values of the phases $\arg t_{ij} = k\pi/4$ ($k\in \mathbb{Z})$  can indeed lead to unwanted additional degeneracies and should therefore be avoided. In presence of finite chemical potential, however, the phases of the coupling strengths are not constant but vary with the control parameter $L_{ij}$, see Eq.~\eqref{eq:coupling}. Values of $|\mu|\gtrsim \pi(4(L_{max}-L_{min})^{-1}$ make it impossible to prevent the unwanted phases from appearing in at least one coupling strength throughout the process. As we will show in the following, braiding protocols based on different loops handle the appearance of these unwanted phases in different ways. For generic loops such as $\Gamma_{I}$, a finite chemical potential dramatically challenges the possibility to reach the adiabatic limit within reasonable braiding times. In contrast, the loop $\Gamma_{II}$ can be shown to mitigate the effects of the chemical potential, making it possible to reach adiabaticity faster and easier.

\subsection{Numerical simulation}

\newcommand{\hsigma}{\hat{\sigma}}
\newcommand{\htau}{\hat{\tau}}

We use the lattice Hamiltonian (\ref{eq:braiding1a}) and simulate its time evolution using the QuTiP package \cite{johansson2012,johansson2013} during a total braiding time $T$. As parameters, we choose $L_{min}=0$ and $L_{max}=10$. Due to the exponential suppression of $|t_{ij}|$, this choice ensures that two parafermions separated by a length $L_{max}$ are effectively decoupled.

We implement the braiding by varying the lengths $L_{ij}$ from $L_{min}$ to $L_{max}$ (or vice versa) with a quadratic time dependence. Between two stages of the braiding protocol, we ramp the lengths up and down, respectively, between two times $\tau_i$ and $\tau_f$ as
\begin{align}
	L_{\uparrow,\tau_i,\tau_f}(\tau) &= L_{min}+L_{max}\left(\frac{\tau-\tau_i}{\tau_f-\tau_i}\right)^2,\\
	L_{\downarrow,\tau_i,\tau_f}(\tau) &=L_{min}+ L_{max}\left(\frac{\tau_f-\tau}{\tau_f-\tau_i}\right)^2.
\end{align}
For the simulation, it is convenient to express the braiding Hamiltonian~(\ref{eq:braiding1a}) as a $\mathbb{Z}_4$ clock model using the Fradkin-Kadanoff transformation for $j = 1, 2$ \cite{fendley12},
\begin{equation}
\chi_{2j-1} = \left(\prod_{k=1}^{j-1}\htau_k\right)\hsigma_j,\qquad \chi_{2j} = \zeta^{3/2}\left(\prod_{k=1}^{j-1}\htau_k\right)\hsigma_j\tau_j
\end{equation}
with $\zeta = e^{i \pi /2}$. At every lattice site, the clock operators $\hsigma$ and the flip operators $\htau$ generalize the Pauli matrices of the $\mathbb{Z}_2$ Ising model, and are given by
\begin{equation}\label{eq:fradkin}
\hsigma = \begin{pmatrix}
1 & 0 & 0 & 0 \\
0 & \zeta &  0 & 0\\
0 & 0 &\zeta^2 &0 \\
0 & 0 	&	0   & \zeta^3
\end{pmatrix}, \qquad \htau = \begin{pmatrix}
0 & 0 & 0 & 1 \\
1 & 0 & 0 & 0 \\
0 & 1 & 0 & 0 \\
0 & 0 & 1 & 0
\end{pmatrix}.
\end{equation}
The $\mathbb{Z}_4$ charge $\hat{Q}=\htau_1\htau_2$, which shifts all the clock variables by $\zeta$, commutes with the full Hamiltonian. We can thus label the eigenstates of the Hamiltonian with its eigenvalues $\{\hat{Q}\}=\{\zeta,\zeta^2,\zeta^3,\zeta^4=1\}$. For the rest of this section, we denote by $\ket{\psi^0_q(\tau)}$ the instantaneous ground states of the Hamiltonian $H(\tau)$. They satisfy $\hat{Q}\ket{\psi^0_q(\tau)}=\zeta^q\ket{\psi^0_q(\tau)}$ and $H(\tau)\ket{\psi^0_q(\tau)}=E^0(\tau)\ket{\psi^0_q(\tau)}$. We will from here henceforth call $\ket{\psi^0_q(\tau)}$ a parity-$q$ ground state.

\subsection{Protocol A: Simultaneous exchange}

\begin{figure}\label{fig:couplength}
	\includegraphics[scale=0.17]{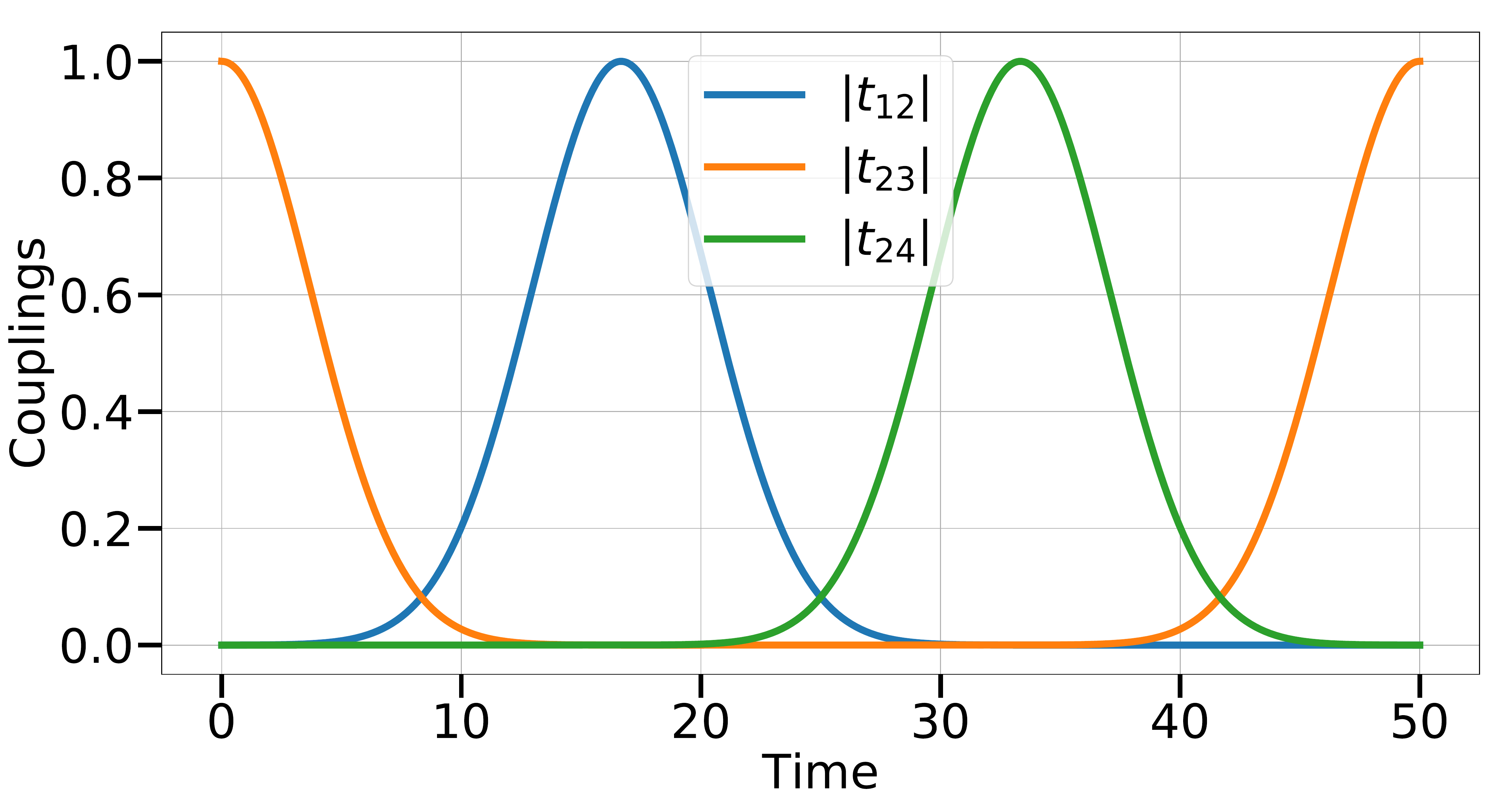}
	\includegraphics[scale=0.17]{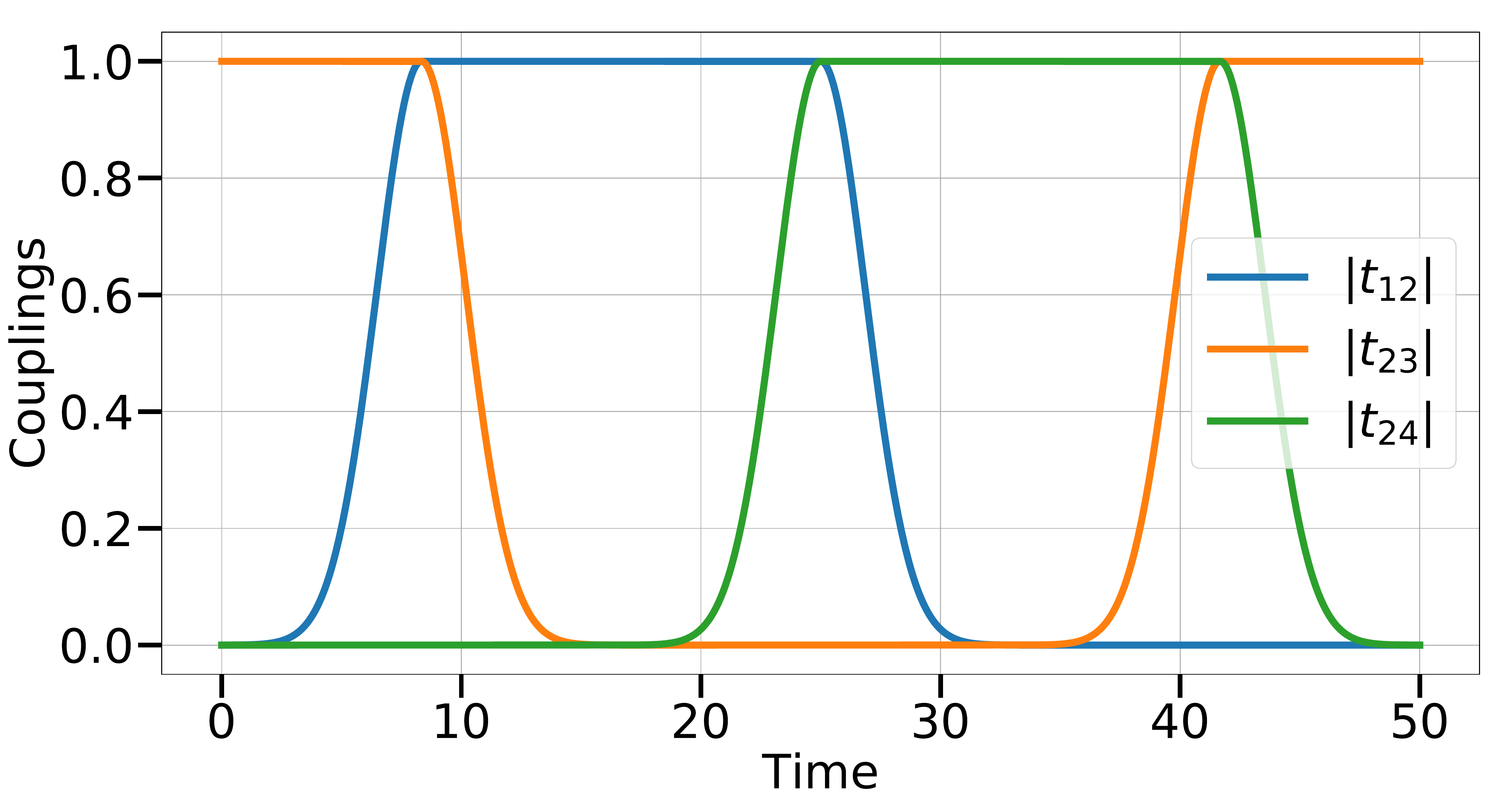}
	\caption{Comparison of the two braiding protocols under consideration. In the simultaneous exchange protocol (upper panel), at most two parafermions are strongly coupled at any given time. For the sequential exchange protocol (lower panel), in contrast, up to three parafermions are strongly coupled.}
	\label{fig:coupling}
\end{figure}
Let us first focus on a braiding protocol which implements a loop like  $\Gamma_I$. As shown in Fig.~\ref{fig:braidingloops}, it features four main steps for times $\tau=0<\tau_1<\tau_2<T$ when the Hamiltonian reads
\begin{align}\label{eq:braidp}
	H(0)&= e^{i\varphi_0}\chi_2\chi_3^\dag + \hc,\\
	H(\tau_1) &= e^{i\varphi_0}\chi_1\chi_2^\dag + \hc,\nonumber\\
	H(\tau_2) &= e^{i\varphi_0}\chi_2\chi_4^\dag + \hc,\nonumber\\
	H(T) &= H(0)\nonumber.
\end{align}
For the numerical solutions, we use a global phase $\varphi_0 = 0.4\pi$, unless specified otherwise, to minimize the effect of accidental degeneracies. The steps of the braiding protocol are depicted in Fig.~\ref{fig:protocolA} and they are characterized by the following values of the lengths
\begin{equation}\label{eq:protocolA}
\begin{tabular}{|c| c c c|}
\hline
$\tau$ &  $L_{23}$ & $L_{12}$ & $L_{24}$ \\
\hline\hline
0  & $0$ & $L_{max}$ & $L_{max}$ \\
$\tau_1$ &  $L_{max}$ & $0$ & $L_{max}$ \\
$\tau_2$ & $L_{max}$ & $L_{max}$ & $0$ \\
$T$ &  $0$ & $L_{max}$ & $L_{max}$ \\
\hline
\end{tabular}
\end{equation}
Between any two stages, two parafermions are moved at the same time. For example, for $0<t<\tau_1$, parafermion $\chi_3$ is moved away from $\chi_2$ while $\chi_1$ is brought closer to $\chi_2$. We therefore refer to such a protocol as the conventional protocol with ``simultaneous exchanges".

\begin{figure}
	\includegraphics[scale=0.18]{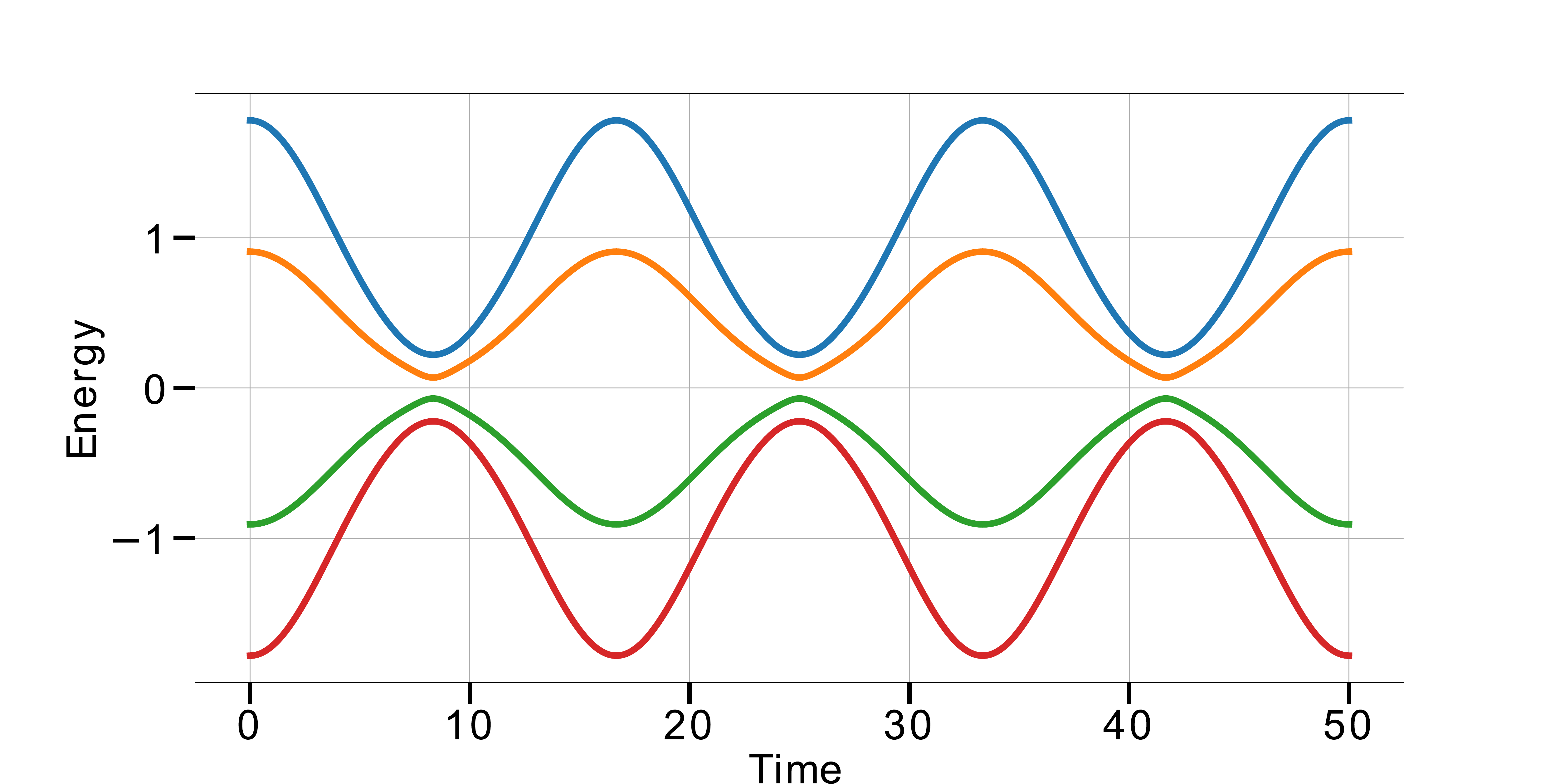}
	\includegraphics[scale=0.18]{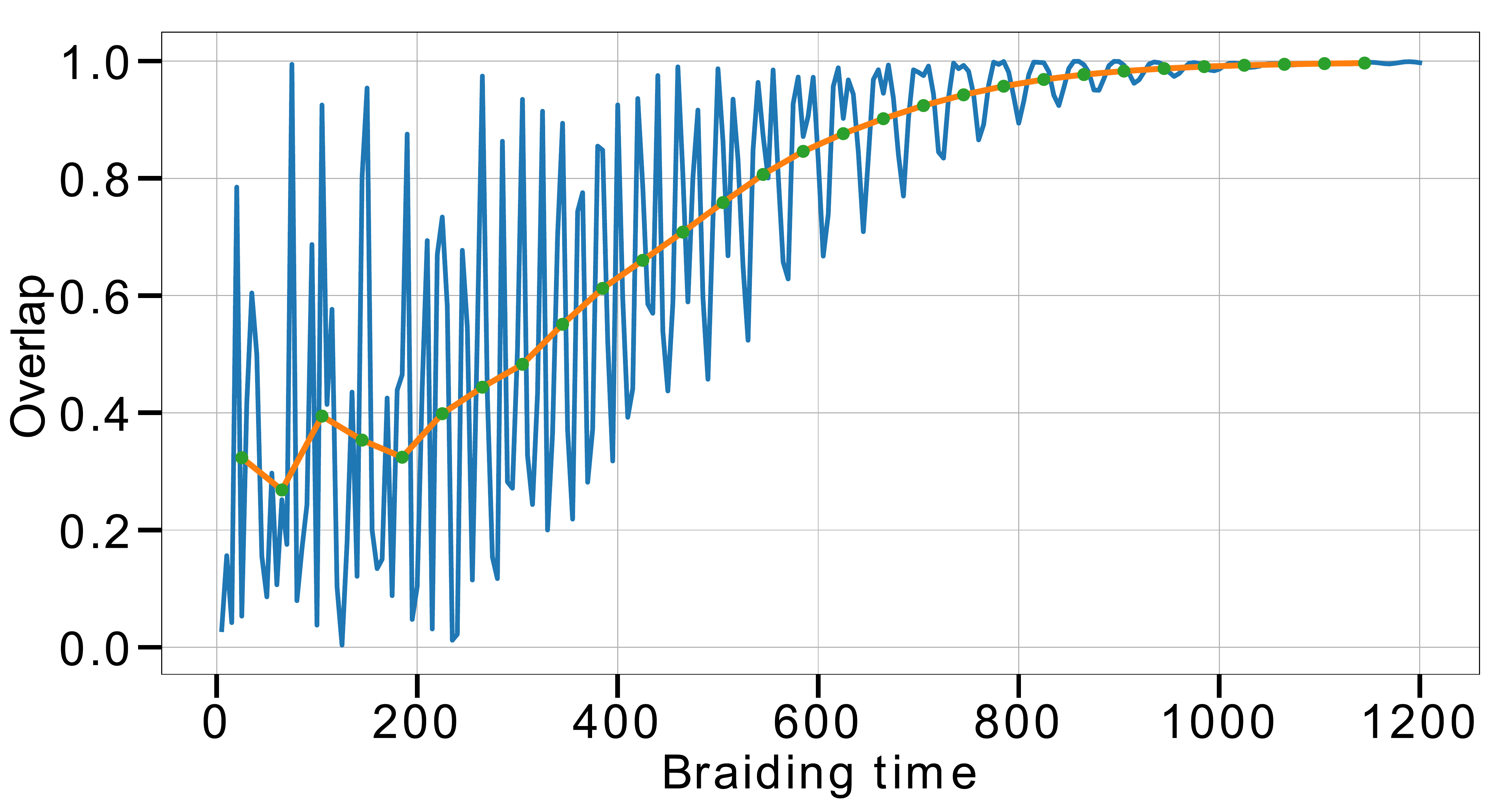}
	\caption{\textit{Upper panel:} Spectrum of the braiding Hamiltonian~(\ref{eq:braiding1a}) for $T=50$ using the simultaneous exchange protocol without chemical potential ($\mu = 0$). \textit{Lower panel:} Overlap $O_0(T)$ for initial state parity $q=0$ as a function of braiding time $T$. The green dots correspond to averaging over 8 different braiding times. The orange curve connects the dots to indicate the general trend of the overlap as a function of braiding time.}
	\label{fig:specFidA}
\end{figure}

In the upper panel of Fig.~\ref{fig:coupling}, we plot the coupling energies as a function of time for a total braiding time $T=50$, and in the upper panel of Fig.~\ref{fig:specFidA} we plot the corresponding instantaneous spectrum of the braiding Hamiltonian~(\ref{eq:braiding1a}). As expected, the spectrum consists of four $4$-fold degenerate manifolds. The energy gaps between successive energy eigenstates are minimal at three times during the braiding process. These points corresponds to the times at midpoints between different stages in the protocol in Eq.~(\ref{eq:protocolA}), i.e., $\tau \in \{ \tau_1/2, (\tau_1 + \tau_2)/2, (\tau_2 + T)/2\}$. At those times, the distances between all parafermions are large, so they become approximately degenerate.

If the braiding is adiabatic, the system follows the instantaneous ground state. In the non-adiabatic case, transitions to excited states can occur. As a figure of merit for braiding protocols, we therefore consider the overlap squared between the (desired) instantaneous parity-$q$ ground state $\ket{\psi^0_q(T)}$ at the final time $T$, and the actual time-evolved state $\ket{\psi(T)}$ of the system prepared initially in the parity-$q$ ground state,
\begin{align}
O_q(T) &= \left|\braket{\psi(T)|\psi^0_q(T)}\right|^2.
\end{align}
The final state $\ket{\psi(T)}$ is calculated by numerically evolving the initial state in time using the time evolution operator,
\begin{align}
\ket{\psi(T)} &= U(T)\ket{\psi^0_q(0)}\notag \\
U(T) &= \mathcal{T}\exp\left[-i\int_0^TH(\tau)d\tau\right],
\end{align}
where $\mathcal{T}$ denotes the time-ordering operator.

The minimal energy gap $\delta$ in the upper panel of Fig.~\ref{fig:specFidA} sets a speed limit for adiabaticity, which can thus be reached for $T \gg  1/\delta$. The gaps are of the order of  $\delta \propto e^{-L/4}$ where $L$ the maximum distance between the parafermions. In the lower panel of Fig.~\ref{fig:specFidA}, we plot the overlaps $O_0(T)$ for zero chemical potential. For our choice of parameters, large oscillations of the figure of merit are visible in the short-time regime $T\lessapprox 1000$. We attribute those to Landau-Zener-St\"uckelberg (LZS) interferences \cite{Shevchenko2010,knapp16}. To support this, we plot in the App.~\ref{Sect:appendixOverlap} the overlaps between the time evolved state of the system and the excited states for a given time during the braiding process. Because the energy gaps are minimal at multiple times during the process, there exist multiple paths in the energy space connecting the ground state at $\tau =0$ to the final state at time $\tau=T$. Interferences of these paths leads to oscillations in the overlap. For $T\gtrapprox 1000$ the overlap $O_q$ tends to unity and adiabaticity is reached.

In App.~\ref{Sect:appendixBerry} we plot the Berry phases for this braiding process and show that they converge to the theoretically expected values~\cite{lindner12}. We also show that inverting the path in configuration space, which amounts to exchanging the zero modes $\chi_1$ and $\chi_4$ counterclockwise rather than clockwise, yields opposite Berry phases as one would expect.

\begin{figure}
	\includegraphics[scale=0.18]{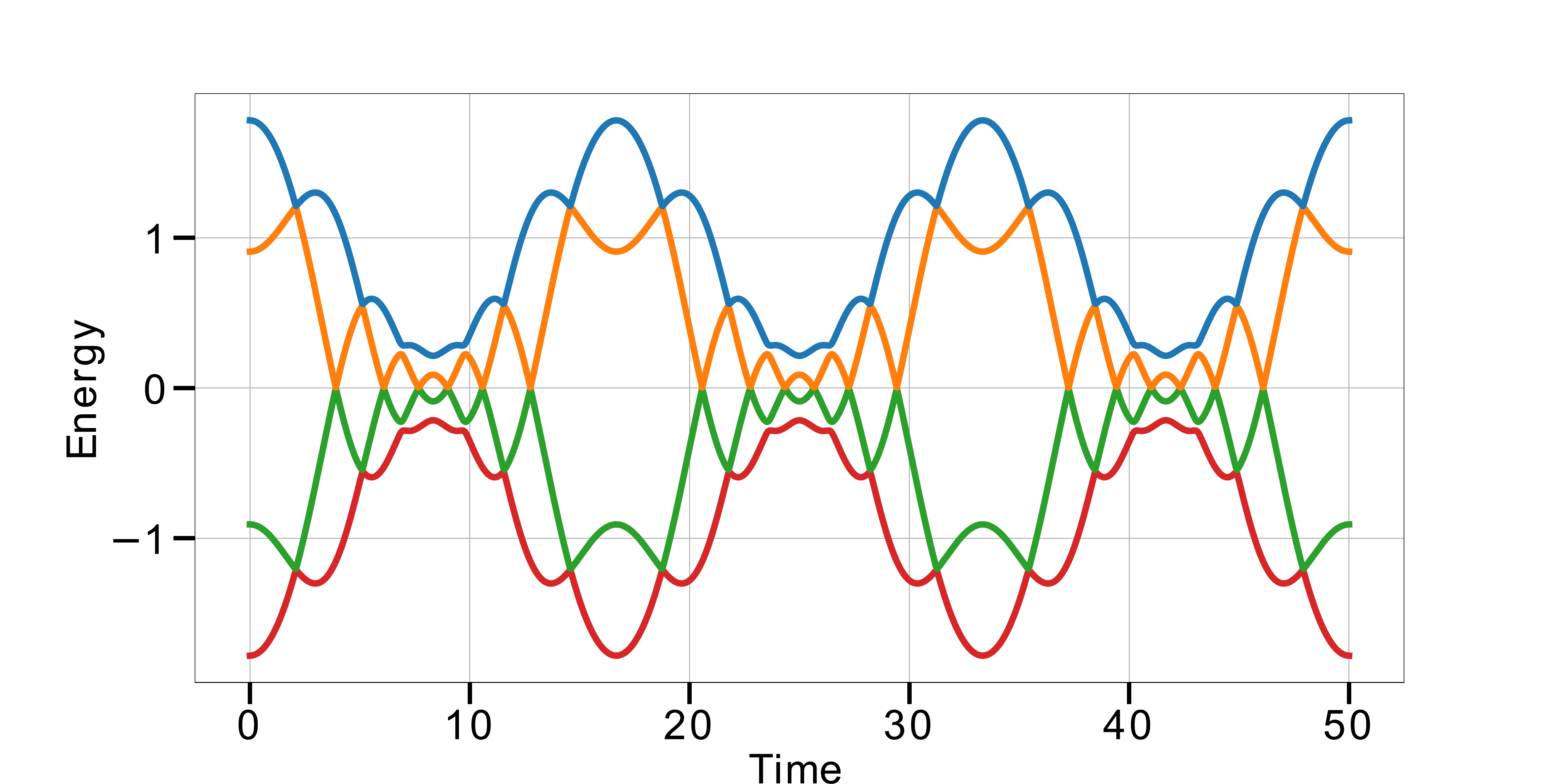}
	\includegraphics[scale=0.18]{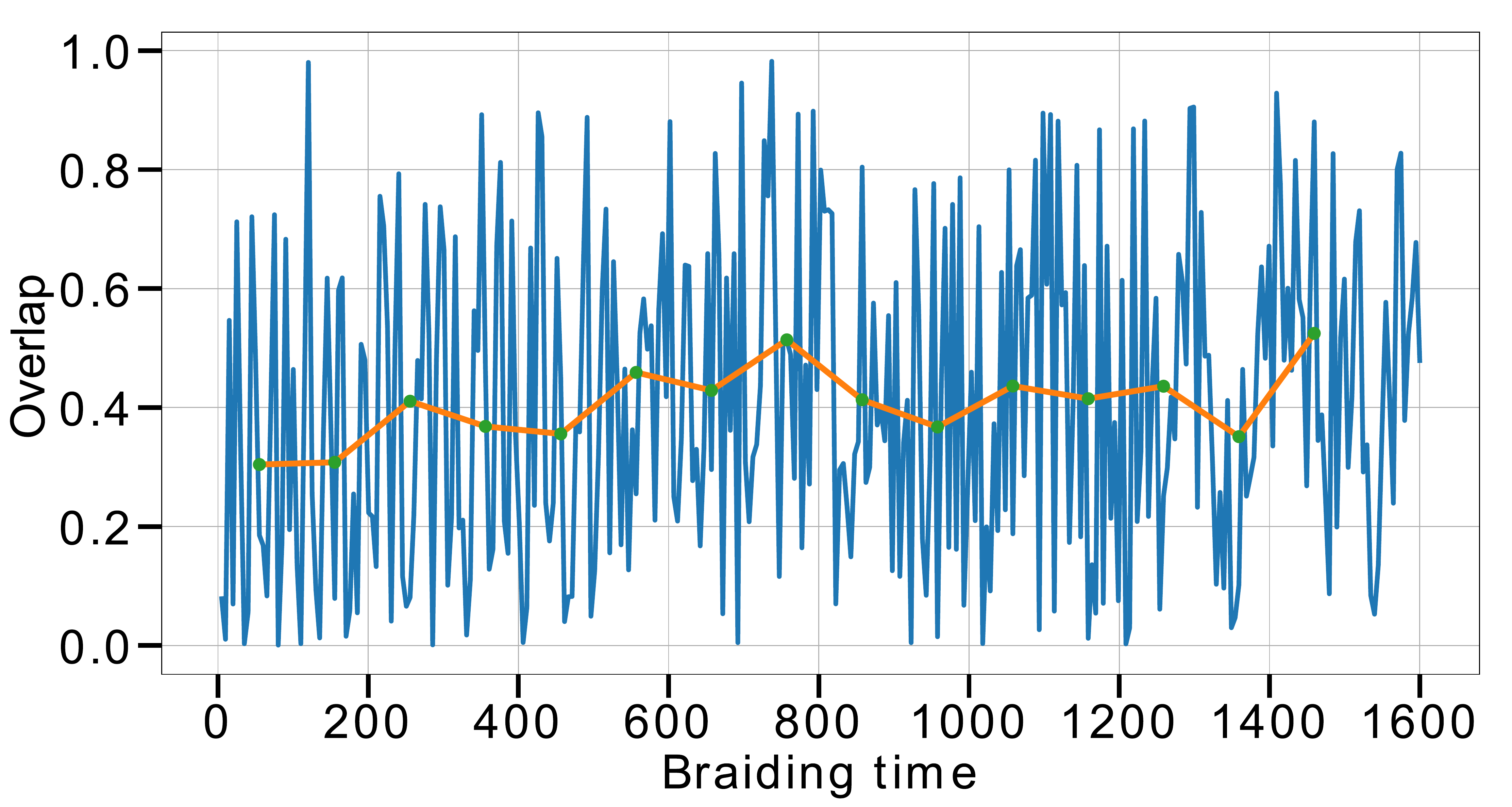}
	\caption{\textit{Upper panel:} Spectrum of the braiding Hamiltonian~(\ref{eq:braiding1a}) for $T=50$ using the simultaneous exchange protocol with nonzero chemical potential $\mu =2$. Exponentially small avoided crossings occur between the energy levels. \textit{Lower panel:} Overlap $O_0(T)$ for initial state parity $q=0$ as a function of total braiding time $T$. Averaging over 20 braiding times is depicted by the green dots, the orange curve indicates the trend. The chemical potential induces many avoided level crossings with exponentially small gaps, so the adiabatic limit is not reached within the simulated time.}
	\label{fig:specFidB}
\end{figure}

For nonzero chemical potential, the scenario becomes more complex. As can be seen from the upper panel of Fig.~\ref{fig:specFidB}, many avoided crossings occur over the entire duration of the braid. The nature of these crossings can be explained by looking at the first stage $\tau \in [0,\tau_{1}]$ of the protocol in Eq.~(\ref{eq:protocolA}). This stage involves a transition from maximal coupling between $\chi_2$ and $\chi_3$ to maximal coupling between $\chi_1$ and $\chi_2$. If we consider the corresponding Hamiltonians $H_{23}(\tau)$ and $H_{12}(\tau)$ separately, the finite chemical potential will induce oscillations in their instantaneous spectra whose amplitudes are controlled by the exponential terms $e^{-L_{23}}$ and $e^{-L_{12}}$. Crossings will then occur in both spectra, where the energy gaps between the respective ground state and first excited state go to zero. As we discuss in App.~\ref{Sect:appendixOrigin}, the full Hamiltonian $H(\tau)$ for $0 \leq \tau \leq \tau_1$ contains both $H_{23}$ and $H_{12}$ contributions, and their combination turns these crossings into avoided crossings. Hence, the energy gap of these avoided crossings is exponentially small in the longer of the distances $L_{23}$ and $L_{12}$.

The presence of these additional avoided crossings has consequences for braiding. As can be seen from the overlap plots in the lower panel of Fig.~\ref{fig:specFidB}, adiabaticity is not reached even at very long times.
A lower bound on the braiding time can then be estimated as $\tau_{\delta(\mu)}= 1/\delta(\mu)$ where $\delta(\mu)$ is the minimal gap over the total braiding time between the ground state and the first excited state as a function of the chemical potential. As explained before, as soon as those additional degeneracies occur, this minimal gap is on the order of $\delta(\mu) \propto e^{-L}$, where $L$ is the maximal distance between the parafermions. It is important to notice that not only the gaps are smaller compared to the case without chemical potential (compared to $\delta\propto e^{-L/4}$), but a chemical potential with $\mu L \gtrapprox 1$ also significantly increases the number of these gaps throughout the spectrum. Since multiple crossings cause LZS interferences, remaining in the ground state requires slower time evolution for an increased number of crossings.

\subsection{Protocol B: Sequential exchange}
\label{subsect:sequential}
\begin{figure}
	\includegraphics[scale=0.45]{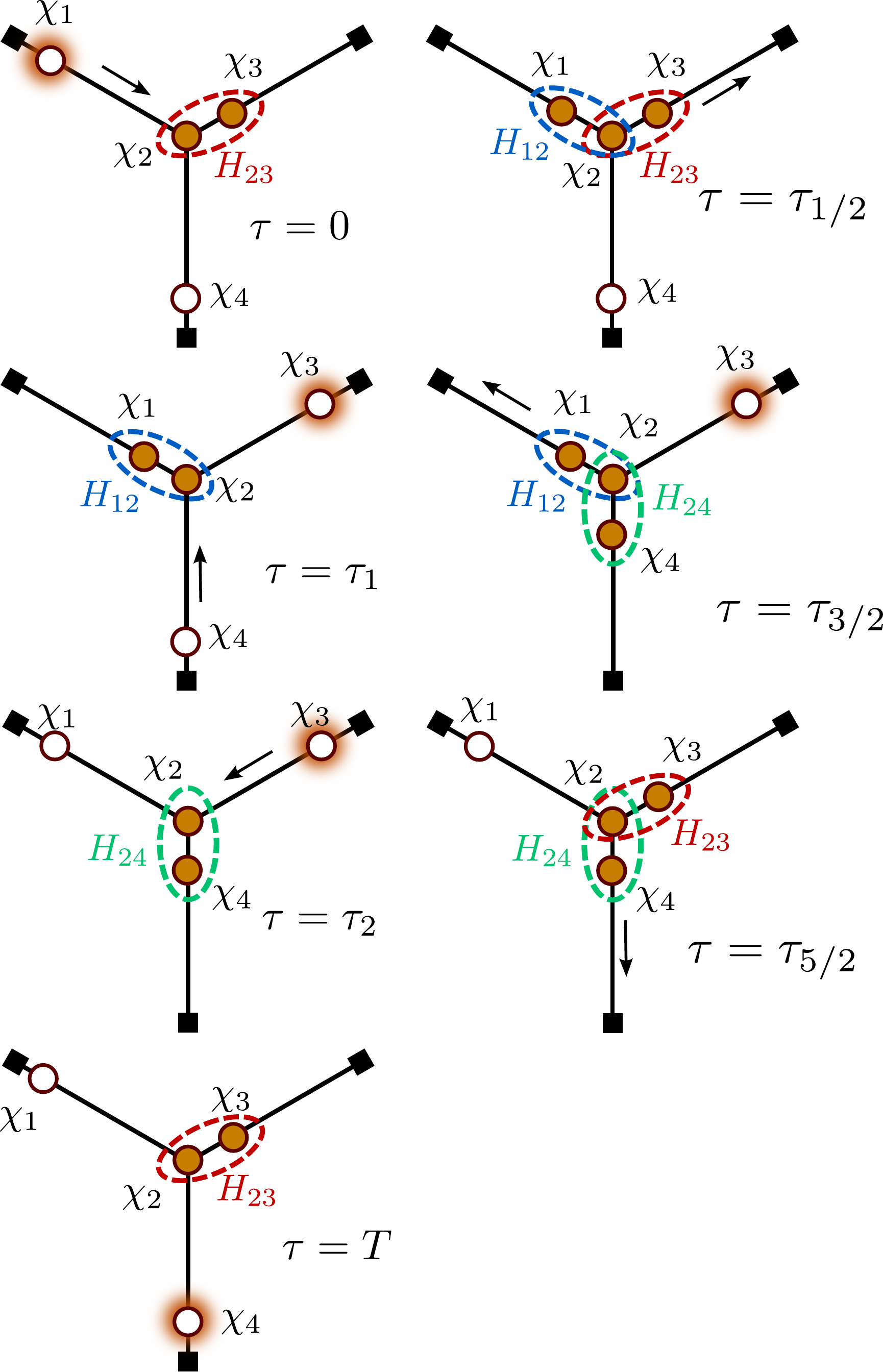}
	\caption{The sequential exchange braiding protocol we propose, for the times $0<\tau_{1/2}<\tau_1<\tau_{3/2}<\tau_{2}<\tau_{5/2}<T$. Empty circles represent zero energy modes and full circles represent finite energy modes. The initial pair $\chi_{2,3}$ stays coupled as one brings $\chi_1$ close to $\chi_2$. This assures that the gap between the ground state manifold and the first excited state stays large. When $\chi_{1,2}$ are maximally coupled, one can then decouple $\chi_{2,3}$. One then continue the steps until $\chi_{2,3}$ are the only coupled parafermionic pairs.}
	\label{fig:protocolB}
\end{figure}

From the insights of the previous subsection, we propose a peculiar braiding protocol which mitigates problems related to adiabaticity and which is based on the $\Gamma_{II}$ loop. As discussed above, the latter can be obtained via a smooth deformation of $\Gamma_{I}$, meaning that they are topologically equivalent and thus implement the same braiding operation. We show this equivalence numerically in App.~\ref{Sect:appendixBerry}. A similar protocol for Majorana zero modes has been investigated by Ref.~\cite{heck12}, and thus our protocol can be seen as a generalization of it to parafermionic modes.
The peculiarity of the new protocol, however, is that at every time at least two parafermions remain fully coupled. This is achieved by adding extra steps in between the ones listed in \eqref{eq:protocolA},
\begin{center}
	\begin{tabular}{|c| c c c|}
		\hline
		$\tau$ &  $L_{23}$ & $L_{12}$ & $L_{24}$ \\
		\hline\hline
		0  & $\mathbf{0}$ & $L_{max}$ & $L_{max}$ \\
		$\tau_{1/2}$ &  $\mathbf{0}$ & $\mathbf{0}$ & $L_{max}$ \\
		$\tau_1$ &  $L_{max}$ & $\mathbf{0}$ & $L_{max}$ \\
		$\tau_{3/2}$ &  $L_{max}$ & $\mathbf{0}$ & $\mathbf{0}$ \\
		$\tau_2$ &  $L_{max}$ & $L_{max}$ & $\mathbf{0}$ \\
		$\tau_{5/2}$ &  $\mathbf{0}$ & $L_{max}$ & $\mathbf{0}$ \\
		$T$ &  $\mathbf{0}$ & $L_{max}$ & $L_{max}$ \\
		\hline
	\end{tabular}
\end{center}
\noindent All of these steps are shown in Fig.~\ref{fig:protocolB}. We refer to this peculiar protocol as ``sequential exchanges protocol'', as only one parafermion is moved at any time. By inspecting for example the first main stage ($0 \leq \tau \leq \tau_1)$, one can indeed see that at first $\chi_1$ is brought closer to $\chi_2$ while keeping $\chi_2$ coupled to $\chi_3$. The latter is then moved away only at later time ($\tau>\tau_{1/2}$).

\begin{figure}
	\includegraphics[width=\columnwidth]{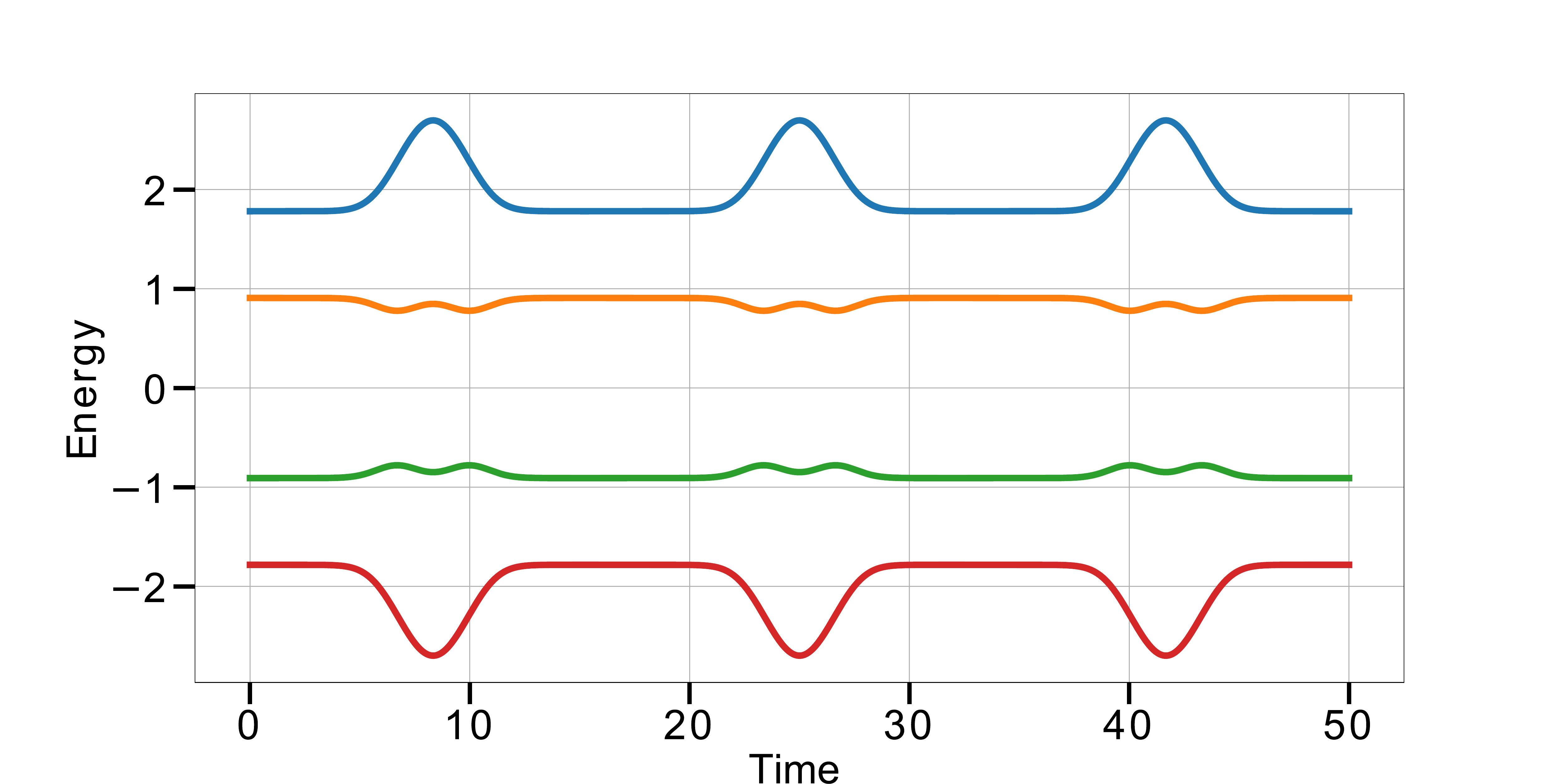}
	\includegraphics[width=\columnwidth]{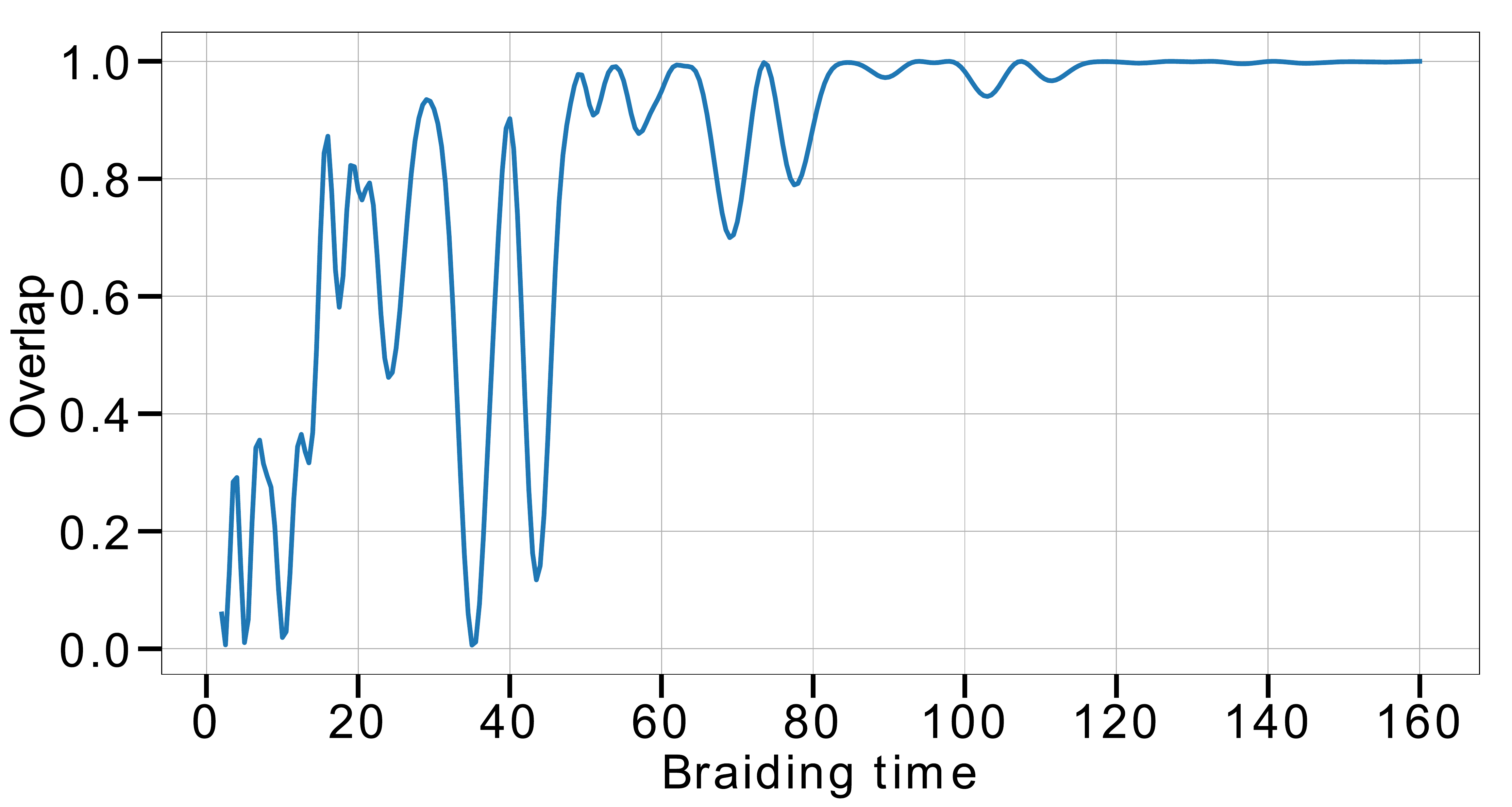}
	\caption{\textit{Upper panel:} Spectrum of the braiding Hamiltonian~(\ref{eq:braiding1a}) for $T=50$ without chemical potential ($\mu = 0$) using the sequential braiding scheme. \textit{Lower panel:} Overlap $O_0(T)$. We see that adiabaticity is reached much faster compared to the simultaneous exchange protocol.}
	\label{fig:specFidAcalzona}
\end{figure}

\begin{figure}
	\includegraphics[width=\columnwidth]{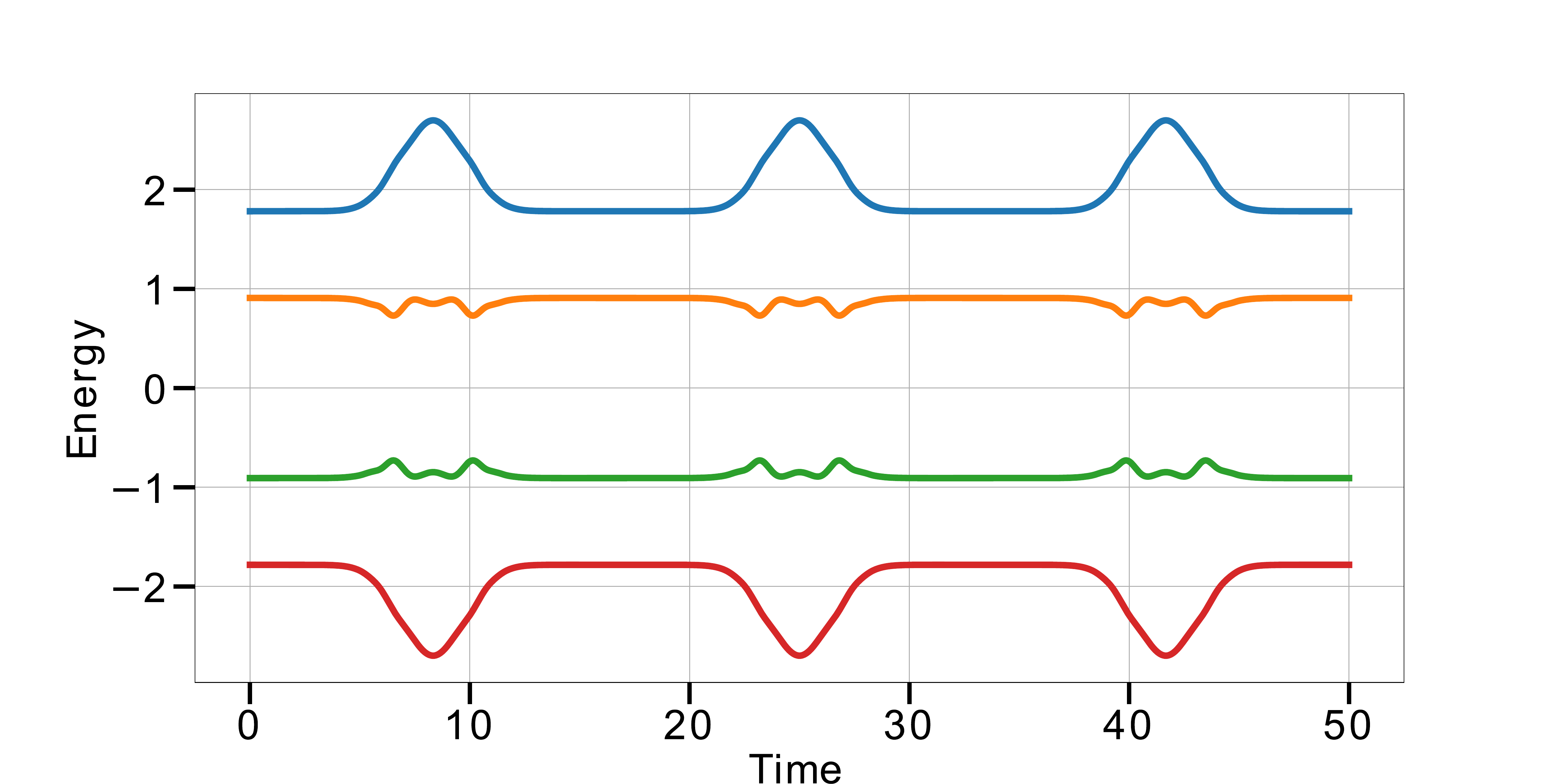}
	\includegraphics[width=\columnwidth]{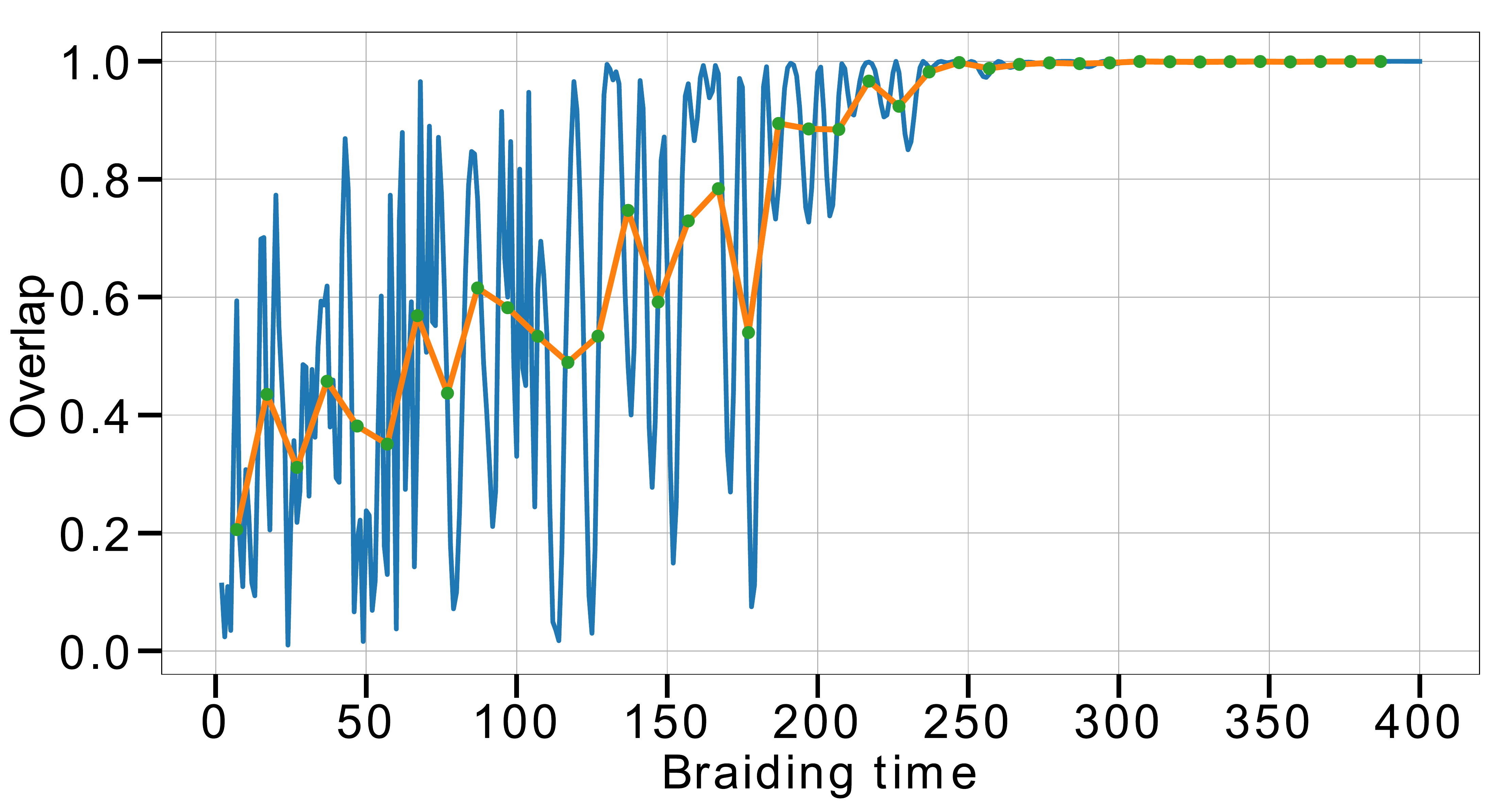}
	\caption{\textit{Upper panel:} Spectrum of the braiding Hamiltonian~(\ref{eq:braiding1a}) for $T=50$ with maximum length $L=10$ using the sequential braiding scheme with $\mu=2$. \textit{Lower panel:} Overlap $O_0(T)$ for every time step and running average over $10$ time steps. The adiabatic limit is reached for much smaller braiding times compared to the simultaneous exchange protocol.}
	\label{fig:specFidAcalzonaB}
\end{figure}

Keeping one parafermionic pair fully coupled makes the gap between the ground state manifold and the excited states as large as possible, as shown in Figs.~\ref{fig:specFidAcalzona} and \ref{fig:specFidAcalzonaB} (upper panels). This makes it possible to reach the adiabatic limit for shorter braiding times with respect to the conventional protocol. This can be seen by comparing the lower panels of Fig.~\ref{fig:specFidAcalzona} and Fig.~\ref{fig:specFidA} where we plot the overlap $O_0(T)$ for the two protocols.

Even for a non-vanishing chemical potential, these features largely persist for the sequential braiding protocol, as can be seen in Fig.~\ref{fig:specFidAcalzonaB}. The adiabatic limit is still reached for much smaller time scales compared to the conventional protocol. The sequential protocol thus permits much faster braiding operations compared to the simultaneous braiding protocol, operating even for finite chemical potentials, when the coupling phases for accidental degeneracies cannot be avoided. Note that the improvements associated with our sequential protocol do not rely on the introduction of extra terms in the Hamiltonian~(\ref{eq:braiding1a}), differently from what discussed in Ref.~\cite{karzig15}.

\section{Conclusion}\label{sec:conclusion}

In this article, we have studied the energy splitting of parafermions separated by a finite distance for parafermions realized either in fractional quantum Hall systems or interacting edge states of two-dimensional topological insulators. We found that in addition to the expected exponential decay with the distance, the energy splitting generally also features an oscillatory term if the gapped region between the parafermions is at a nonzero chemical potential. This property is analogous to a similar result for Majorana bound states and confirms a recent result for parafermions based on a semiclassical approximation.

We use the energy splitting to deduce the complex hopping amplitudes between parafermions in a 1D chain and use this model to perform a numerical simulation of the complete braiding process. Using the conventional braiding protocol, we find that a finite chemical potential induces level crossings at isolated points in the braiding process which make it very hard to reach an adiabatic limit. Therefore, we have proposed a modified braiding protocol, which offers larger gaps at all intermediate stages even in the presence of a nonzero chemical potential. We have argued that an adiabatic limit is easier to reach in this improved protocol.

\begin{acknowledgments}
The authors acknowledge support by the National Research Fund, Luxembourg under grants ATTRACT 7556175, INTER 11223315, AFR 11224060 and PRIDE/15/10935404. AC acknowledges support by the W\"urzburg-Dresden Cluster of Excellence on Complexity and Topology in Quantum Matter (EXC 2147, project-id 39085490).
\end{acknowledgments}

\bibliography{references}

\clearpage
\onecolumngrid

\appendix
\section{Overlap between time evolved wave function and excited states}
\label{Sect:appendixOverlap}

\begin{figure}
\includegraphics[scale=0.16]{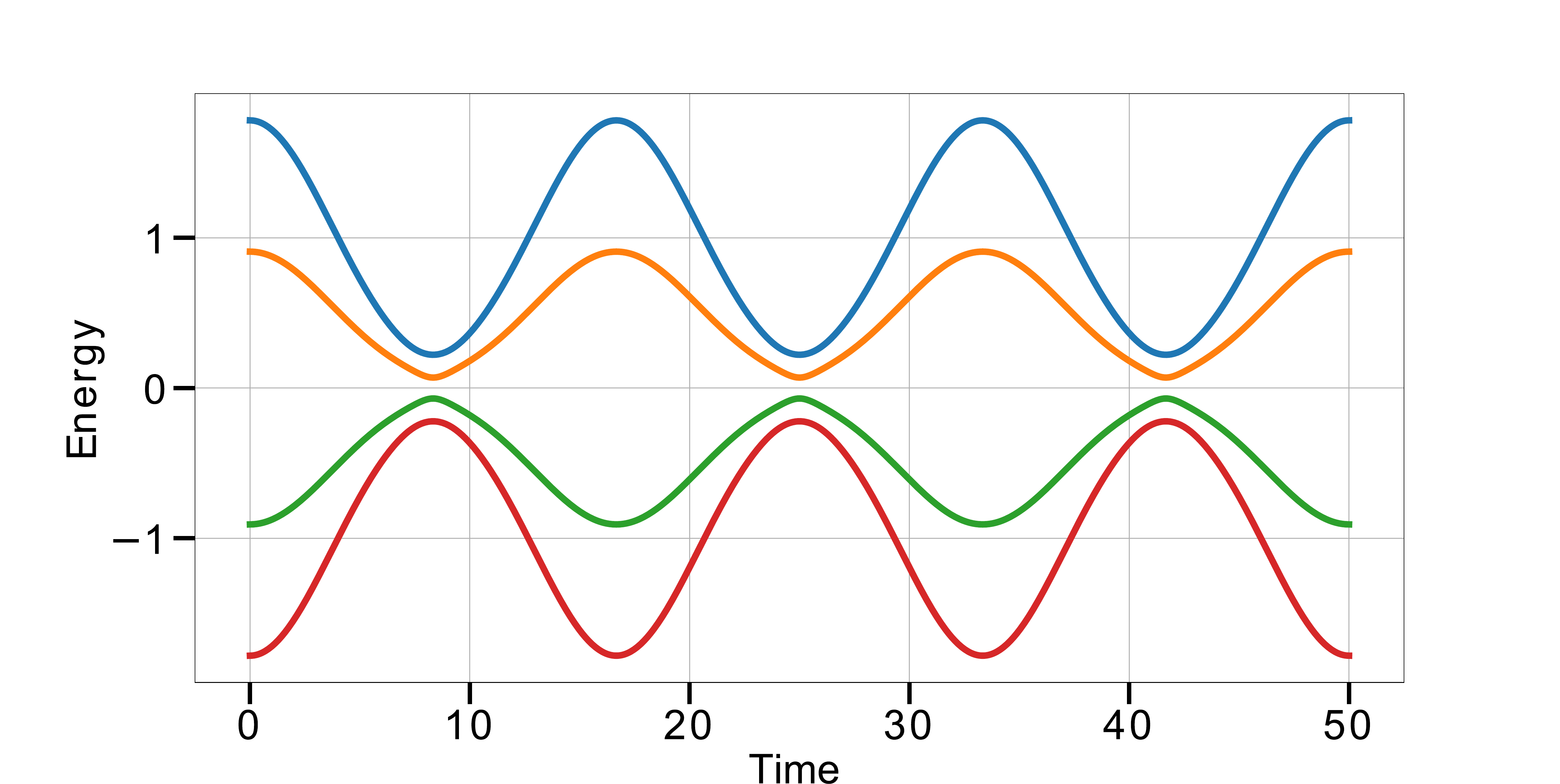}
\includegraphics[scale=0.16]{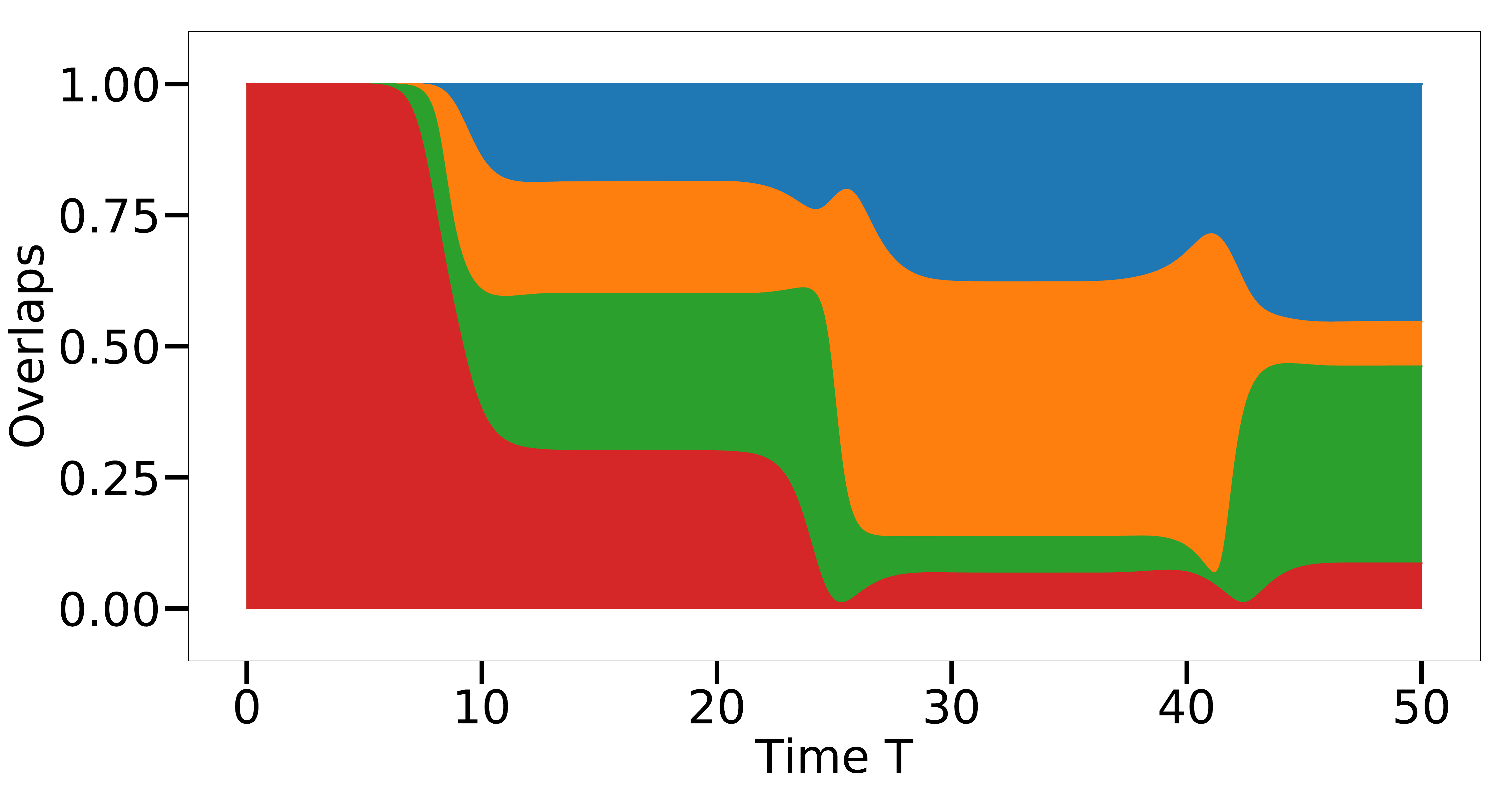}
\includegraphics[scale=0.16]{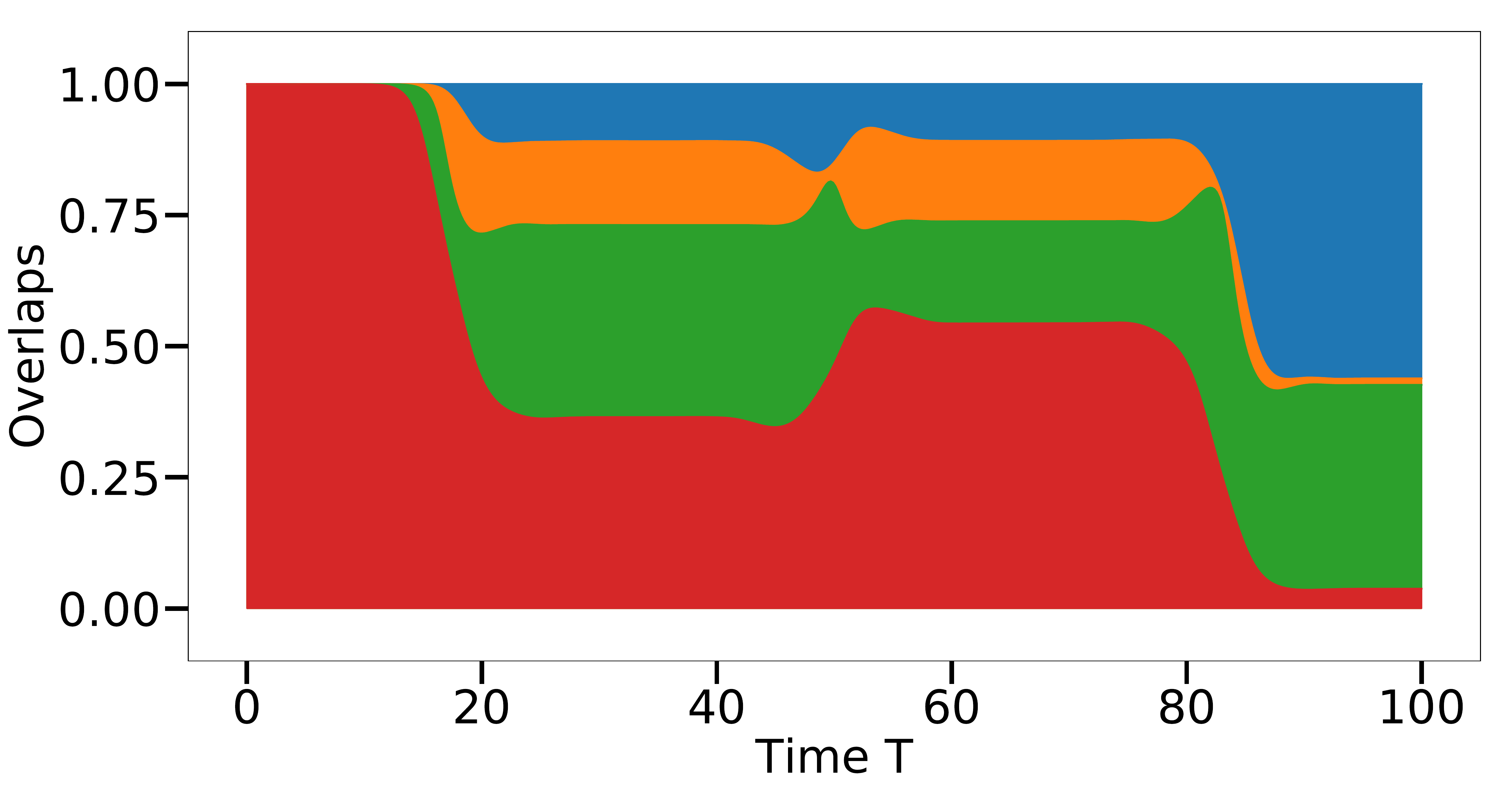}
\includegraphics[scale=0.16]{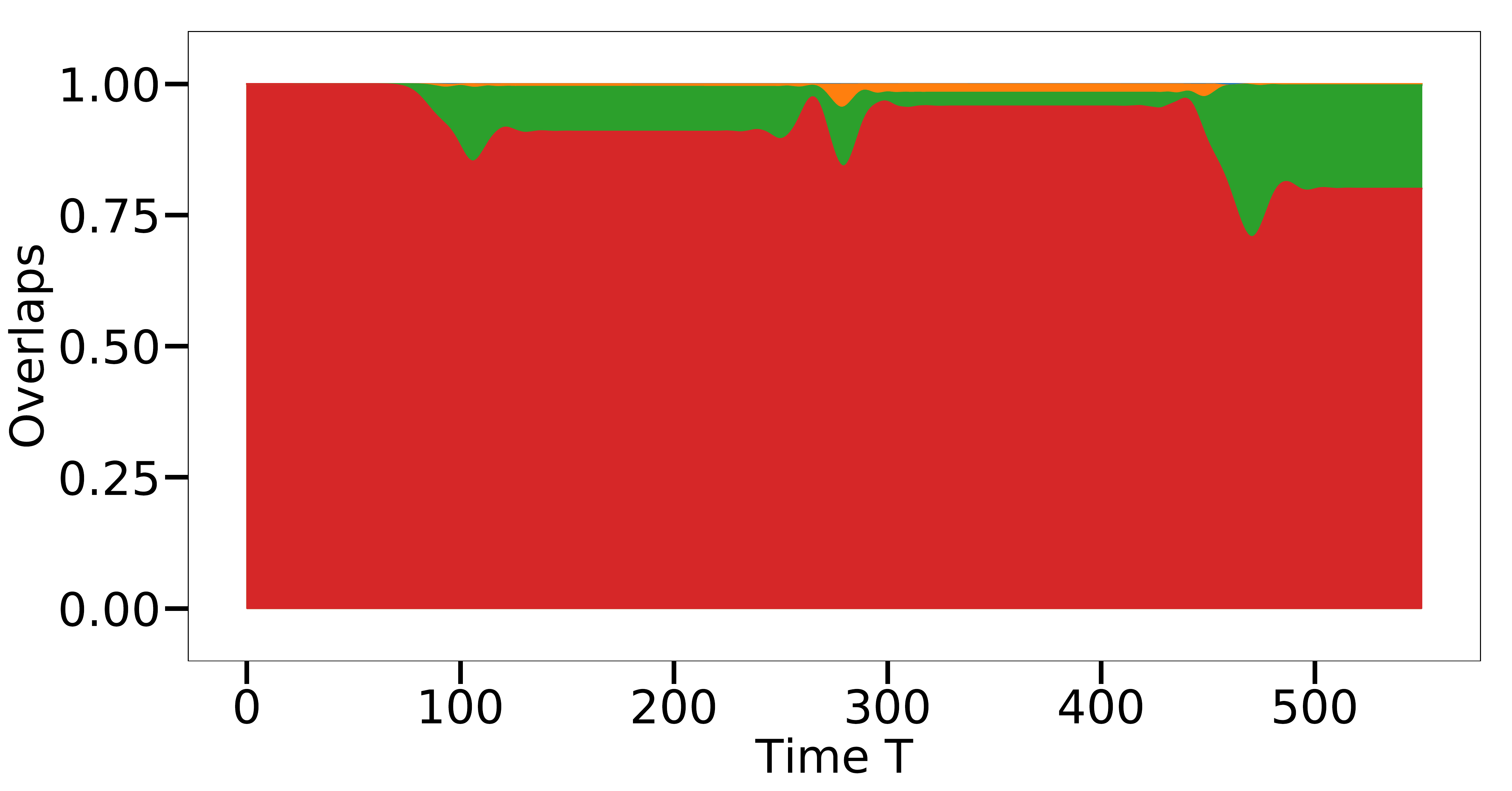}
\caption{\textit{Upper left panel:} Spectrum of the Hamiltonian~(\ref{eq:braiding1a}) for a total braiding time $T=50$ using the simultaneous exchange protocol. \textit{Upper right and lower panels:} The probabilities $V_0^j(\tau)$ to end up in the different states for different total braiding times $T=50$, $T=100$, and $T=550$, using the parameters $\mu = 0$ and global phase $\varphi_0 = 0.4\pi$. The red part indicates the probability to stay in the ground state, other colors represent the excited states.}
\label{fig:appendixA}
\end{figure}

\begin{figure}
\includegraphics[scale=0.16]{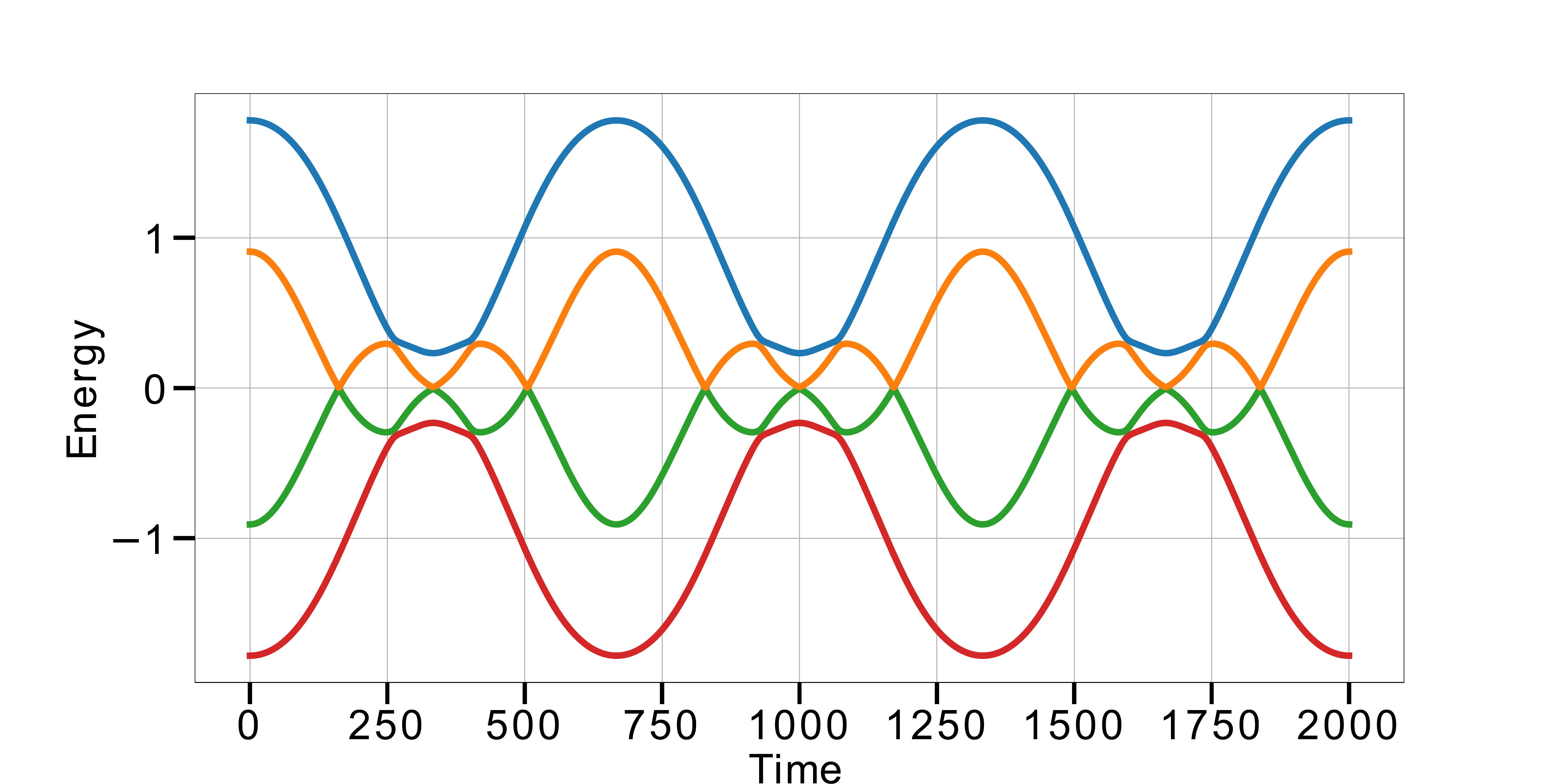}
\includegraphics[scale=0.16]{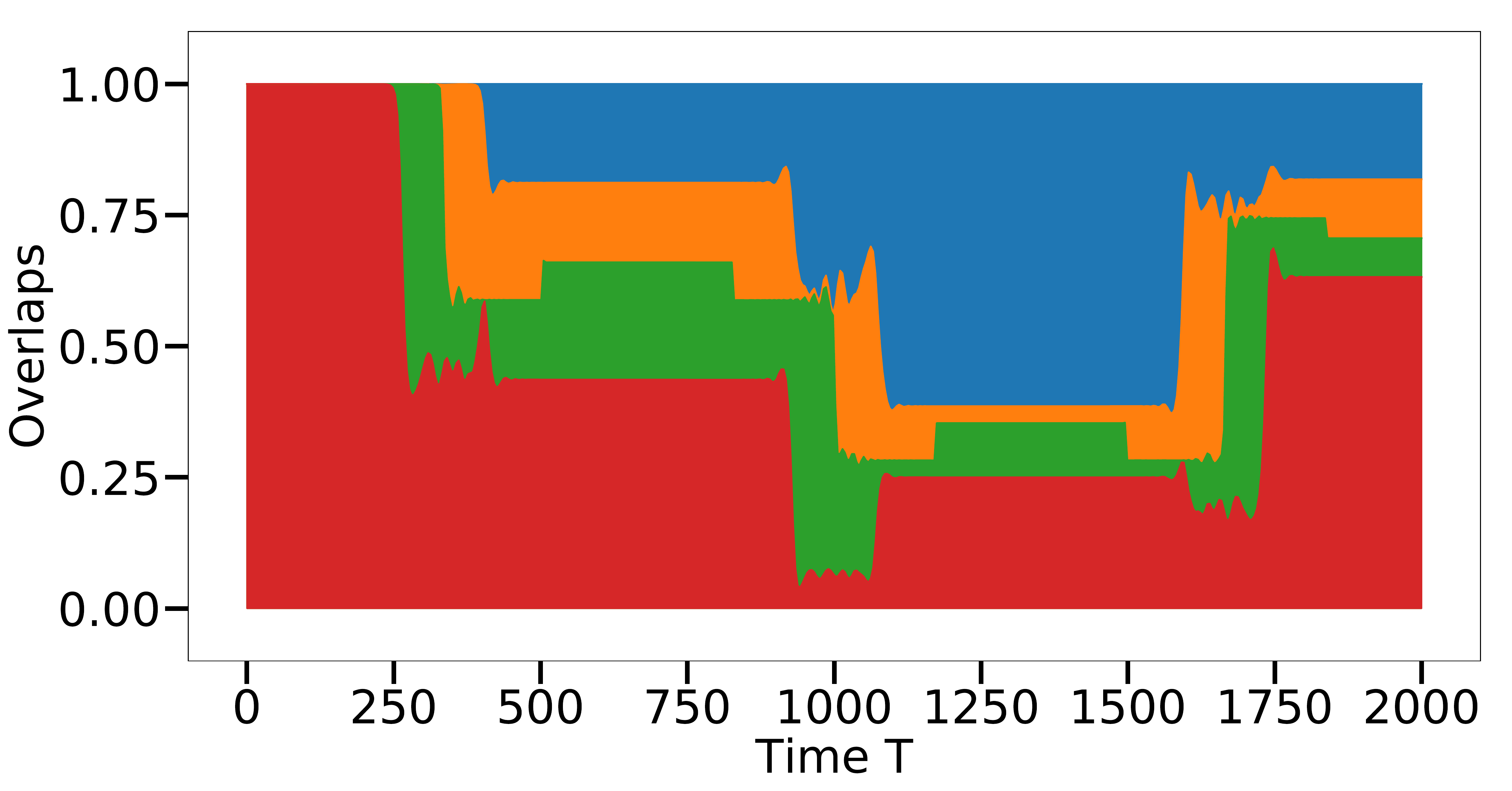}
\includegraphics[scale=0.16]{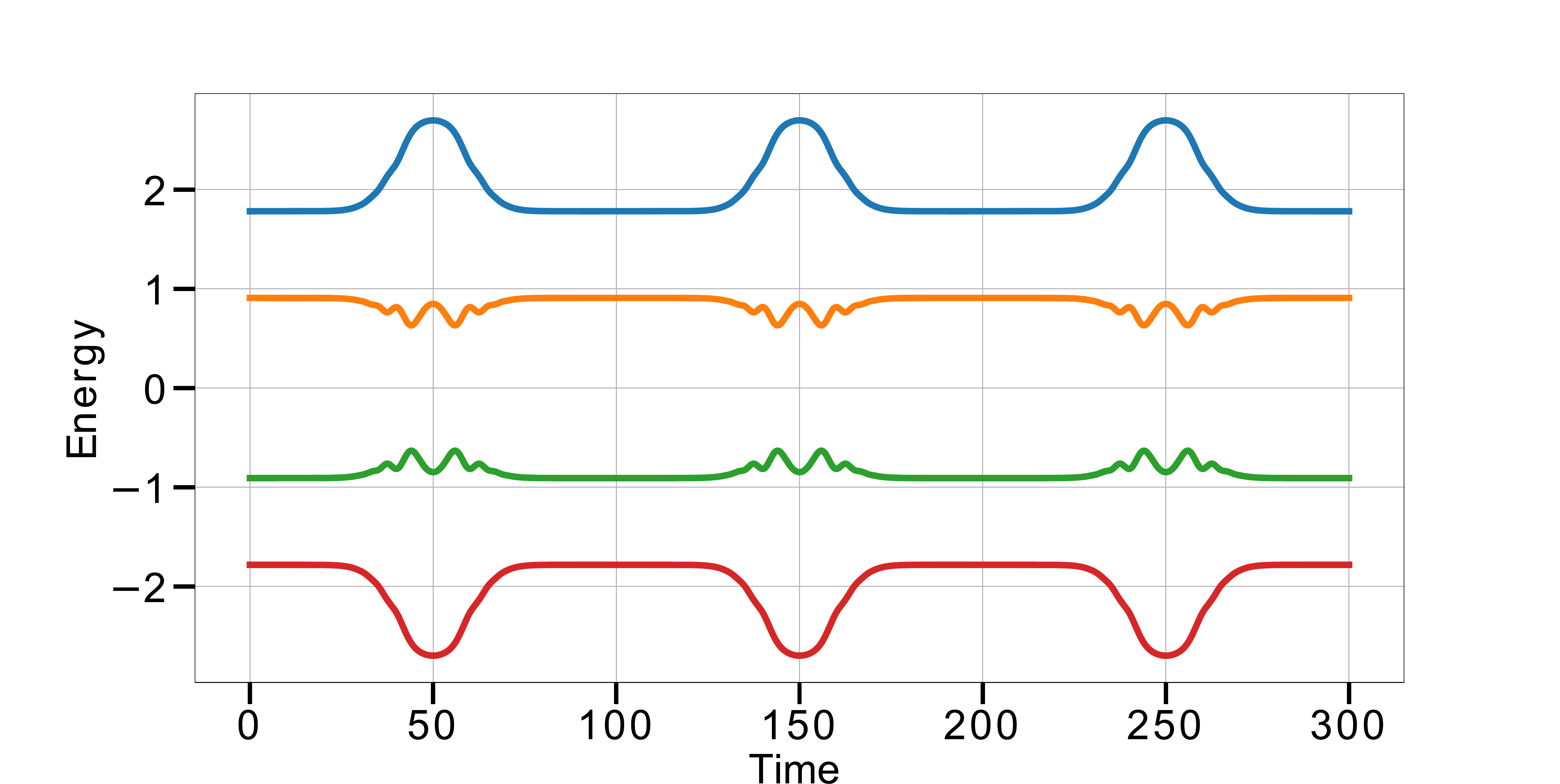}
\includegraphics[scale=0.16]{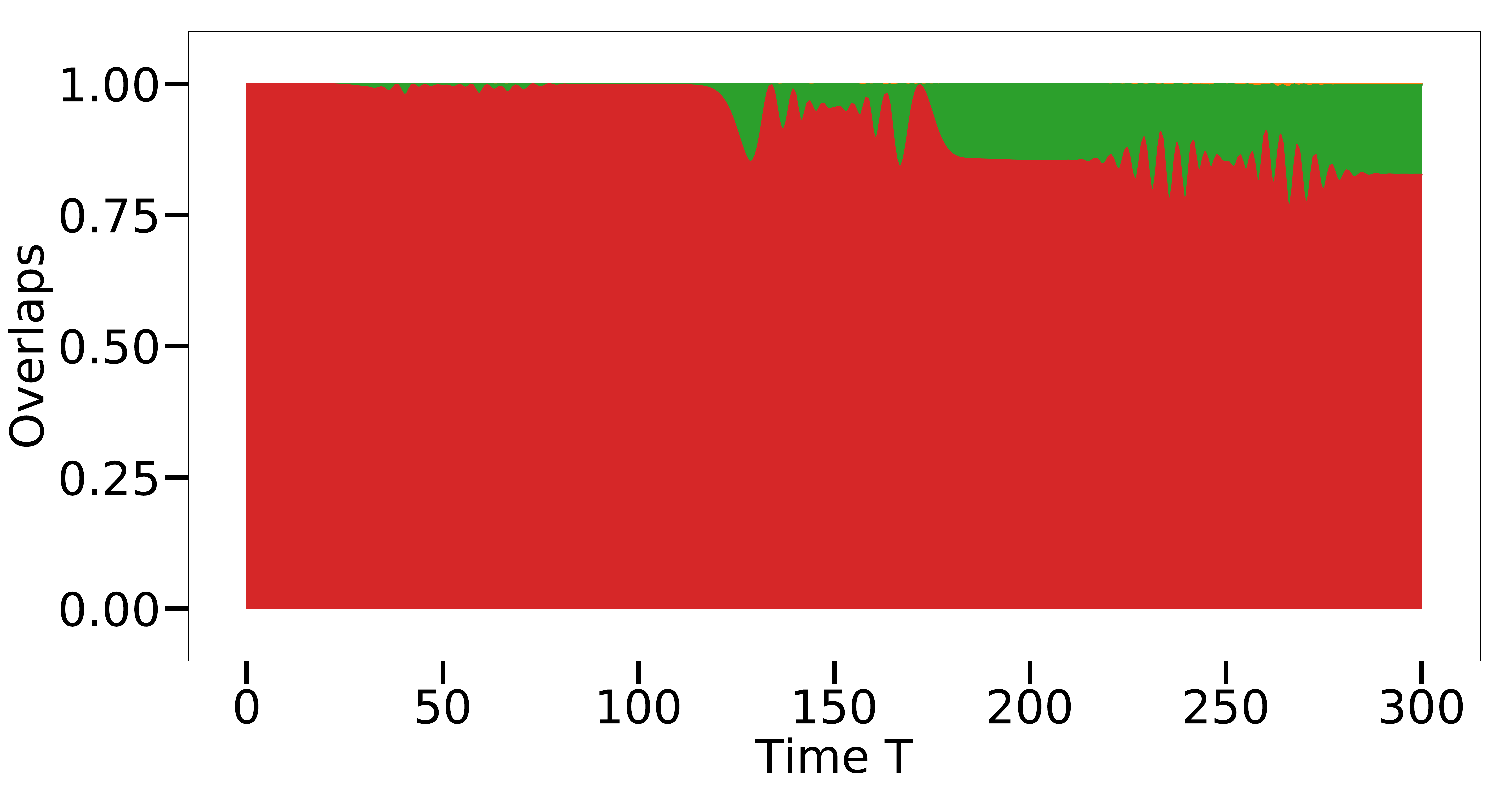}
\caption{Spectrum and overlaps for finite chemical potential, with global phase $\varphi_0 = 0.1\pi$. \textit{Upper panels:} Chemical potential $\mu=0.8$ and simultaneous exchange protocol. \textit{Lower panels:} Chemical potential $\mu=3$ and sequential exchange protocol. In the simultaneous exchange protocol, multiple non-adiabatic transitions occur and destroy the overlap even for relatively low chemical potential and long braiding times. A much higher overlap can be obtained for shorter braiding times and higher chemical potential using the improved sequential exchange protocol.}
\label{fig:appendix2A}
\end{figure}

For any finite braiding time, non-adiabatic transitions outside the ground state manifold can occur. These can be quantified by computing the overlap between the time-evolved wavefunction of the system $\ket
{\psi(\tau)}=U(\tau)\ket{\psi_q^{0}(0)}$ and the instantaneous energy eigenstates $\ket{\psi_q^j(\tau)}$ ($j=0,1,2,3$) for a given parity $q$.
We define the probability to end up in the $j^{th}$ excited state at time $\tau$ by
\begin{equation}
V^{(j)}_q(\tau) = |\braket{\psi(\tau)|\psi_q^{j}(\tau)}|^2.
\end{equation}
As can be seen in the Fig.~\ref{fig:appendixA}, non-adiabatic transitions occur every time the gaps reach a minimum. As these involve multiple levels, and occur multiple times throughout the braiding process, adiabaticity is lost.
These transitions can be suppressed by increasing the total braiding time as can be seen from the lower panels of Fig.~\ref{fig:appendixA}. This scenario persists for non-zero chemical potential, where because of multiple avoided crossings in the spectrum, the adiabatic regime is only reached for much longer braiding times. As can be seen from Fig.~\ref{fig:appendix2A}, sequentially exchanging the parafermions permits much faster braiding operations.

\section{Berry phase}

\label{Sect:appendixBerry}
\begin{figure}
\includegraphics[scale=0.19]{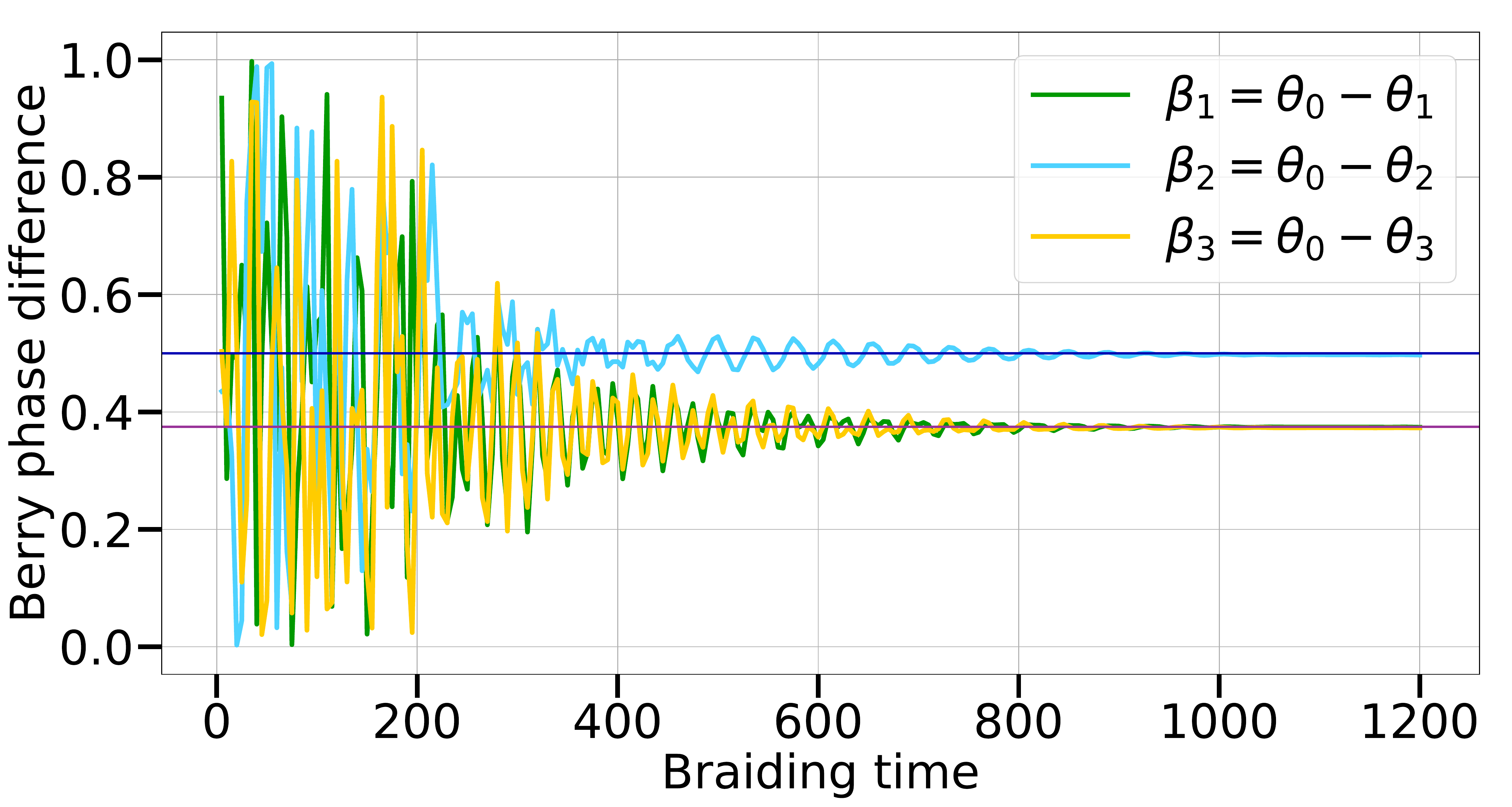}\includegraphics[scale=0.19]{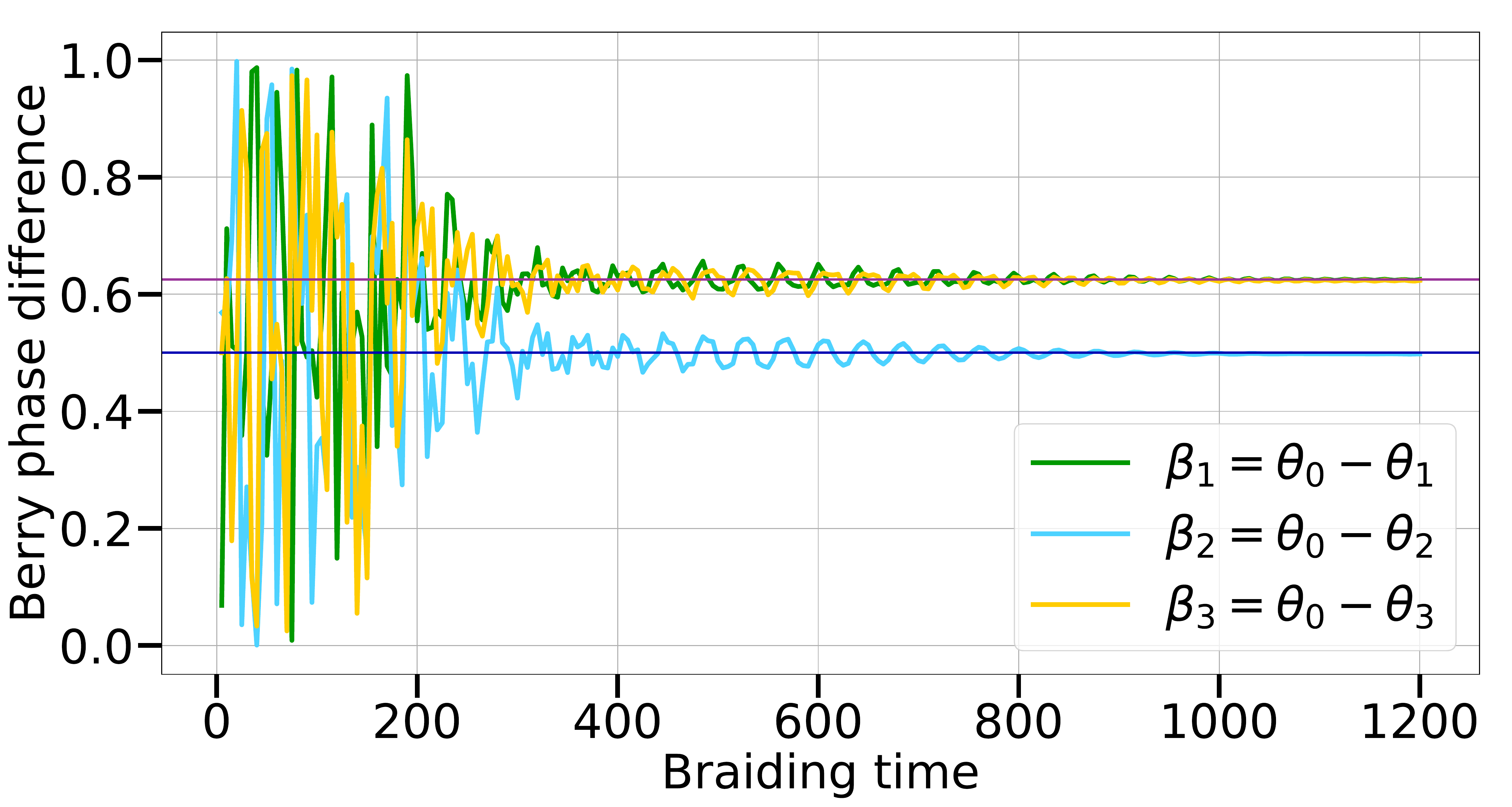}
\caption{The Berry phase difference for $\mathbb{Z}_4$ parafermions for the simultaneous exchange protocol (left panel), and for the simultaneous exchange protocol with inverted path in parameter space (right panel). The phase differences are plotted in units of $2\pi$. \textit{Left panel:} The blue line at $1/2$ indicates that the Berry phase $\beta_2 \to \pi$, whereas the purple line at $3/8$ shows that the Berry phase $\beta_{1,3} \to 3\pi/4$. \textit{Right panel:} For the inverted path, $\beta_2 \to \pi$ whereas $\beta_{1,3} \to 5\pi/4 = -3 \pi/4\ (\text{mod } 2\pi)$. These result are in accordance with the analytical Berry phase in Eq.~(\ref{eq:lindnerberry}). }
\label{fig:Berryz4}
\end{figure}

In the case of an adiabatic exchange, the final and initial state will only differ by a phase,
\begin{align}
\ket{\psi(T)} &= e^{i\phi_q(T)}\ket{\psi_q(0)} =e^{i\theta(T)+i\gamma_q(T)}\ket{\psi_q(0)},
\end{align}
where the dynamical phase and the Berry phase read respectively
\begin{align}
\theta(T) &= \int_0^T E^0(\tau)d\tau \quad\text{and}\quad \gamma_q(T) = i\int_0^T d\tau \bra{\psi_q(\tau)}\partial_\tau\ket{\psi_q(\tau)}.
\end{align}
The former only depends on the instantaneous ground state energy $E^0(\tau)$, whereas the latter explicitly depends on the parity of the ground state. By computing the phase differences between different parity states the dynamical phase will drop out and we obtain thus information about the Berry phase.
Defining $\phi_q=\arg\braket{\psi^0_q(0)|\psi_q(T)}$, we define the Berry phase difference
\begin{equation}
\beta_q = [2\pi+(\phi_0-\phi_q)]\text{ mod }2\pi = (\gamma_0-\gamma_q)\text{ mod }2\pi.
\end{equation}
Adding $2\pi$ and taking the modulo $2\pi$ ensures that $\beta_q\in [0,2\pi]$.
According to Ref.~\cite{lindner12}, the Berry phase which the parity-$q$ state picks up during a full braiding process described by the Hamiltonian~(\ref{eq:braiding1a}) for $\mathbb{Z}_p$ parafermions is given by
\begin{equation}\label{eq:lindnerberry}
\gamma_q = \frac{\pi}{p}(q-k)^2.
\end{equation}
When the complex phases of all coupling strengths are equal, then the integer $k$ is determined by the global phase $\varphi_0$ of the coupling strengths of the Hamiltonian~(\ref{eq:braiding1a}) by
\begin{equation}
\frac{2\pi}{p}k< (\varphi_0+\pi) <\frac{2\pi}{p}(k+1),
\end{equation}
In other words, $k$ is given by the integer part of $ p[\varphi_0+\pi] /(2\pi) $. The difference of $\pi$ with respect to Ref.~\cite{lindner12} stems from a slightly different convention in defining the coupling coefficients.
For the choice of global phase in the main text $\varphi_0 = 0.4\pi$ we have $k=2$ so that $\beta_q = \pi -\gamma_q \text{ mod }2\pi$.
In Fig.~\ref{fig:Berryz4} we plot the Berry phases as a function of total braiding time using the sequential exchange protocol corresponding to a clockwise exchange (left panel) of the zero modes ($\chi_1,\chi_4$) as well as the counterclockwise exchange of the pair (right panel). This is numerically implemented by interchanging $L_{12}\leftrightarrow L_{24}$.
The two scenarios correspond to opposite loops in the control parameter space and give rise to opposite Berry phase differences. Indeed, if $U_{14}$ is a representation of the braid group which exchanges the parafermions $\chi_1$ and $\chi_4$ clockwise then
\begin{equation}
U_{14}\ket{\psi_q(0)} = e^{i\gamma_q}\ket{\psi_q(0)}.
\end{equation}
The operator representing the counterclockwise exchange is represented by $U_{41} =U_{14}^{-1}$.
It thus follows that
\begin{equation}
U_{41}U_{14}\ket{\psi_q(0)} = e^{i\gamma_q}U_{41}\ket{\psi_q(0)} \stackrel{!}{=}\ket{\psi_q(0)},\quad\Rightarrow\quad U_{41}\ket{\psi_q(0)} = e^{-i\gamma_q}\ket{\psi_q(0)}.
\end{equation}
The Berry phases in Fig.~\ref{fig:Berryz4} converge to the theoretical values, and the counterclockwise exchange yields opposite Berry phases.

\begin{figure}[!htbp]
 \includegraphics[scale=0.197]{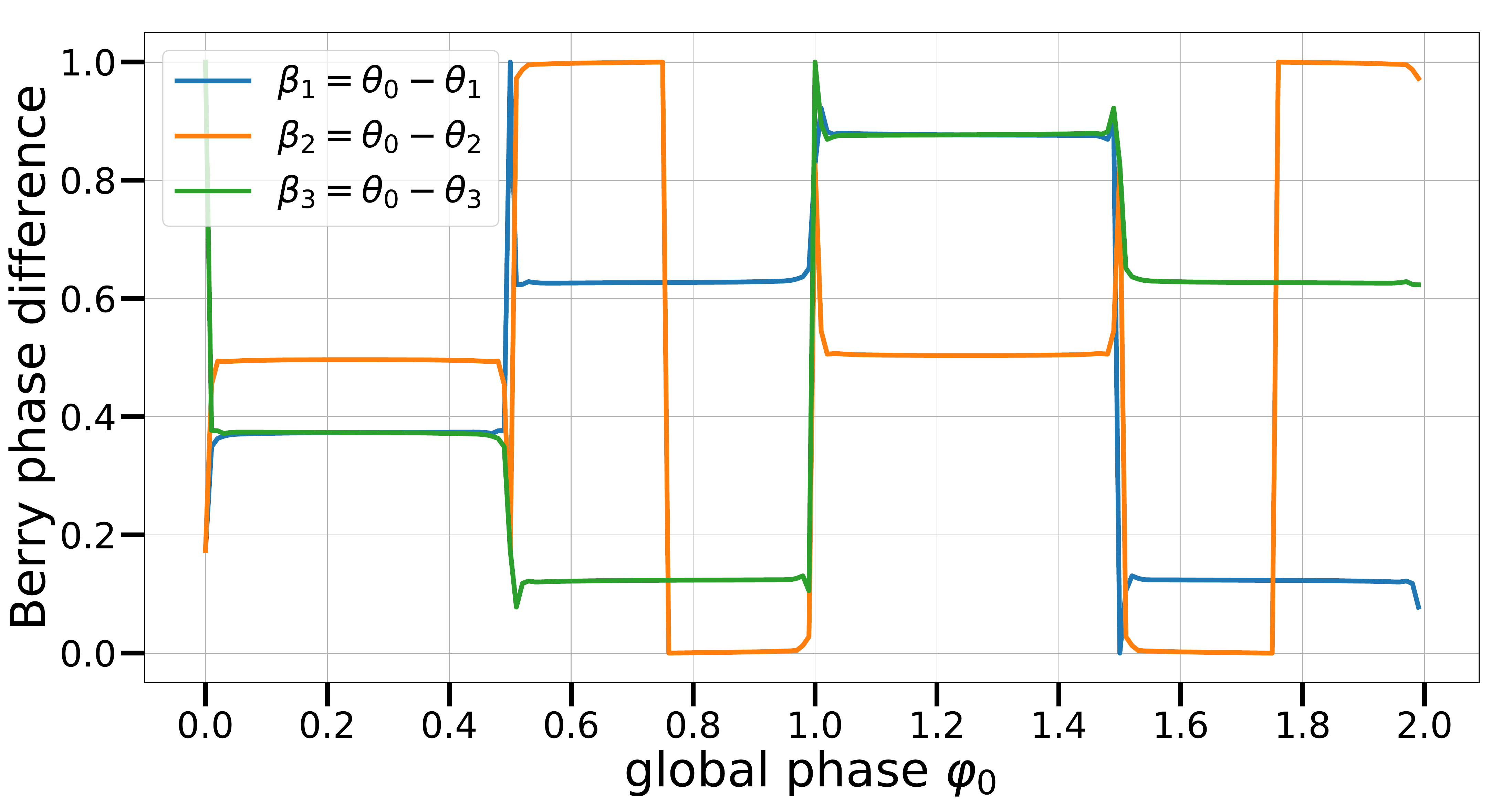}
 \includegraphics[scale=0.197]{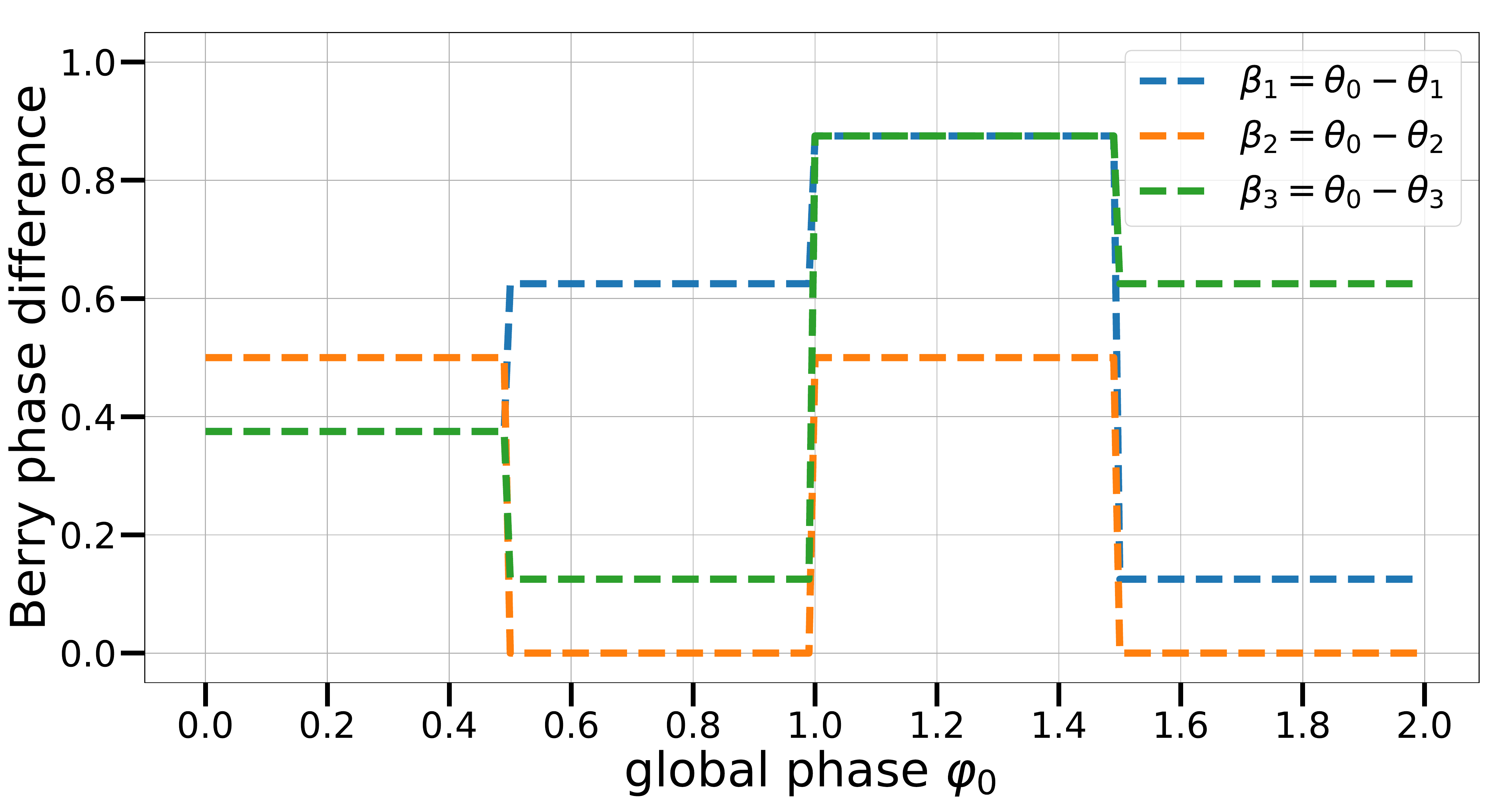}
  \includegraphics[scale=0.197]{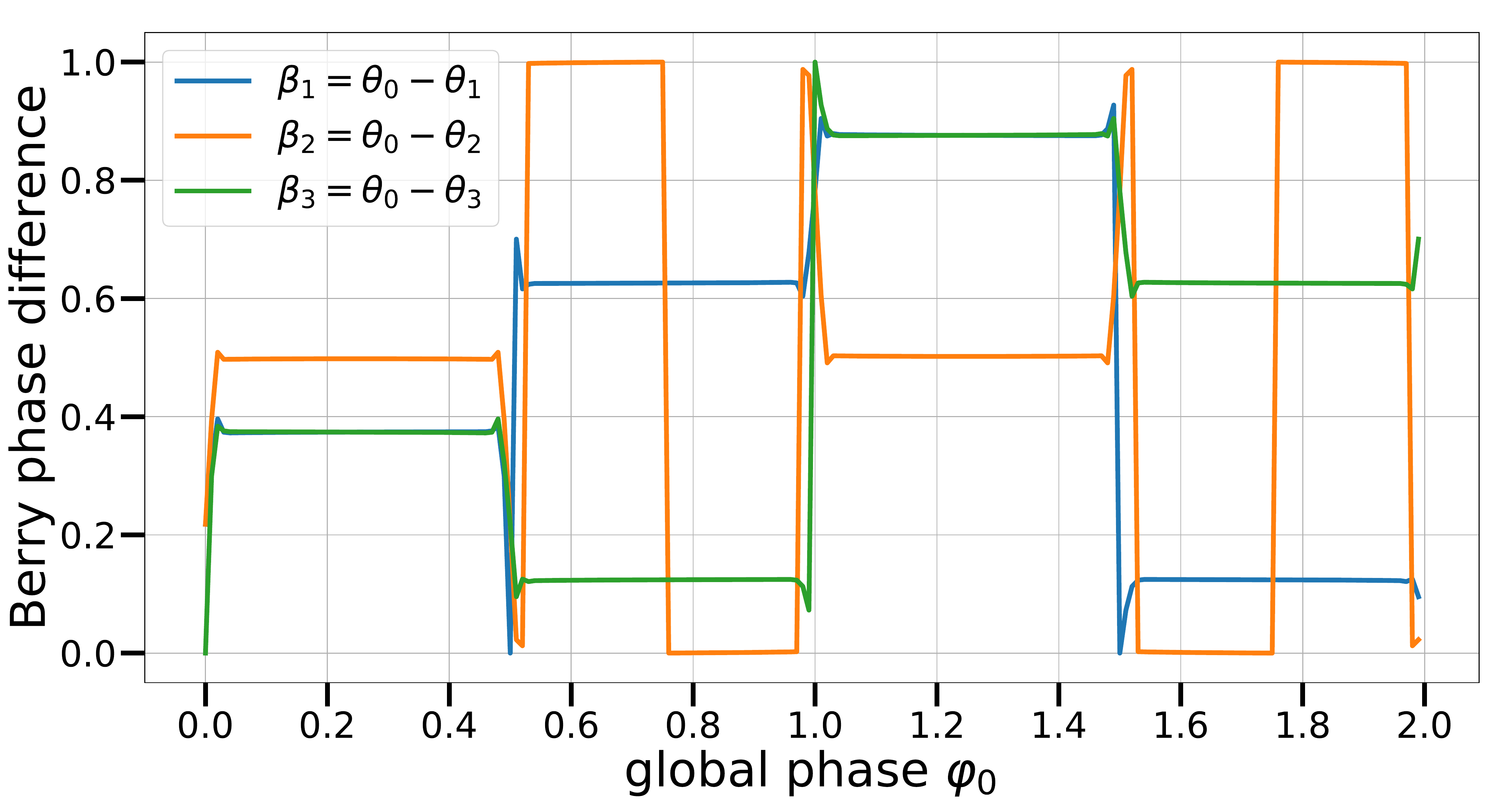}
\caption{\textit{Upper panels:} Berry phase differences as a function of the global phase for $\mathbb{Z}_4$ parafermions using the simultaneous exchange protocol, from numerical simulation (left panel) and using Eq.~(\ref{eq:lindnerberry}) (right panel). \textit{Lower panel:} The Berry phase differences converge to the expected values also using the sequential exchange braiding protocol.}
\label{fig:BerryZ4protocol}
\end{figure}

In Fig.~\ref{fig:BerryZ4protocol} we plot the Berry phase differences as a function of the global phase (left panel), which we compare to the expected Berry phases~(\ref{eq:lindnerberry}) (right panel). The Berry phase differences are computed for $T=2000$ at zero chemical potential. Note that jumps in the Berry phases occur when the global phases crosses the values $m2\pi/p$ for $m\in\mathbb{Z}$: The spectrum becomes degenerate at these points and thus adiabaticity is no longer reached. The Berry phases thus become ill-defined. Jumps between $0$ and $2\pi$ are also to be expected because we are taking the modulo $2\pi$ of the phase differences.

In the lower panel of Fig.~\ref{fig:BerryZ4protocol}, we show that the sequential exchange protocol gives rise to the same Berry phases, and thus ultimately to the same braiding operators as the simultaneous exchange protocol.

\section{Origin of the avoided crossings}
\label{Sect:appendixOrigin}

\begin{figure}
\includegraphics[scale=0.17]{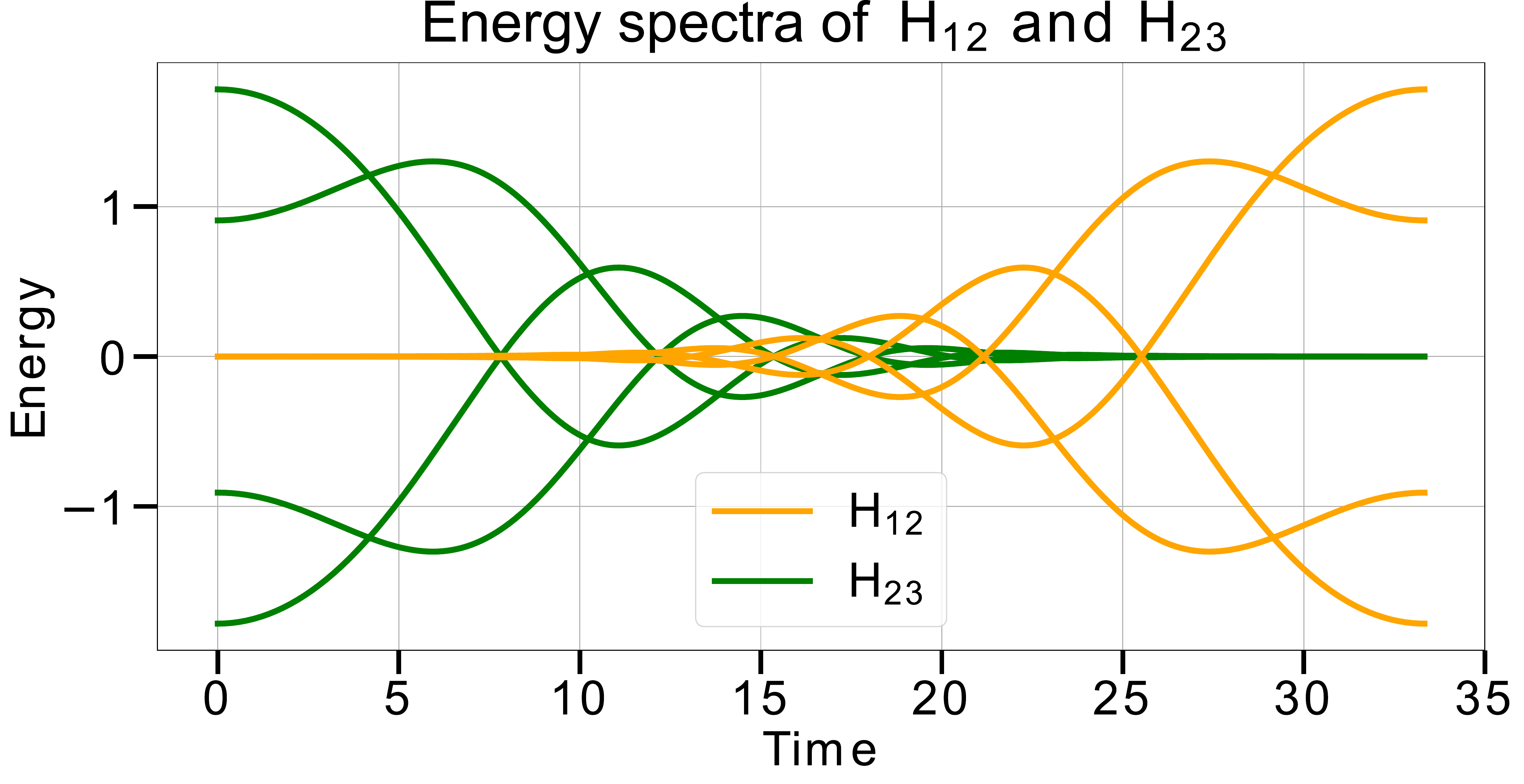}
\includegraphics[scale=0.17]{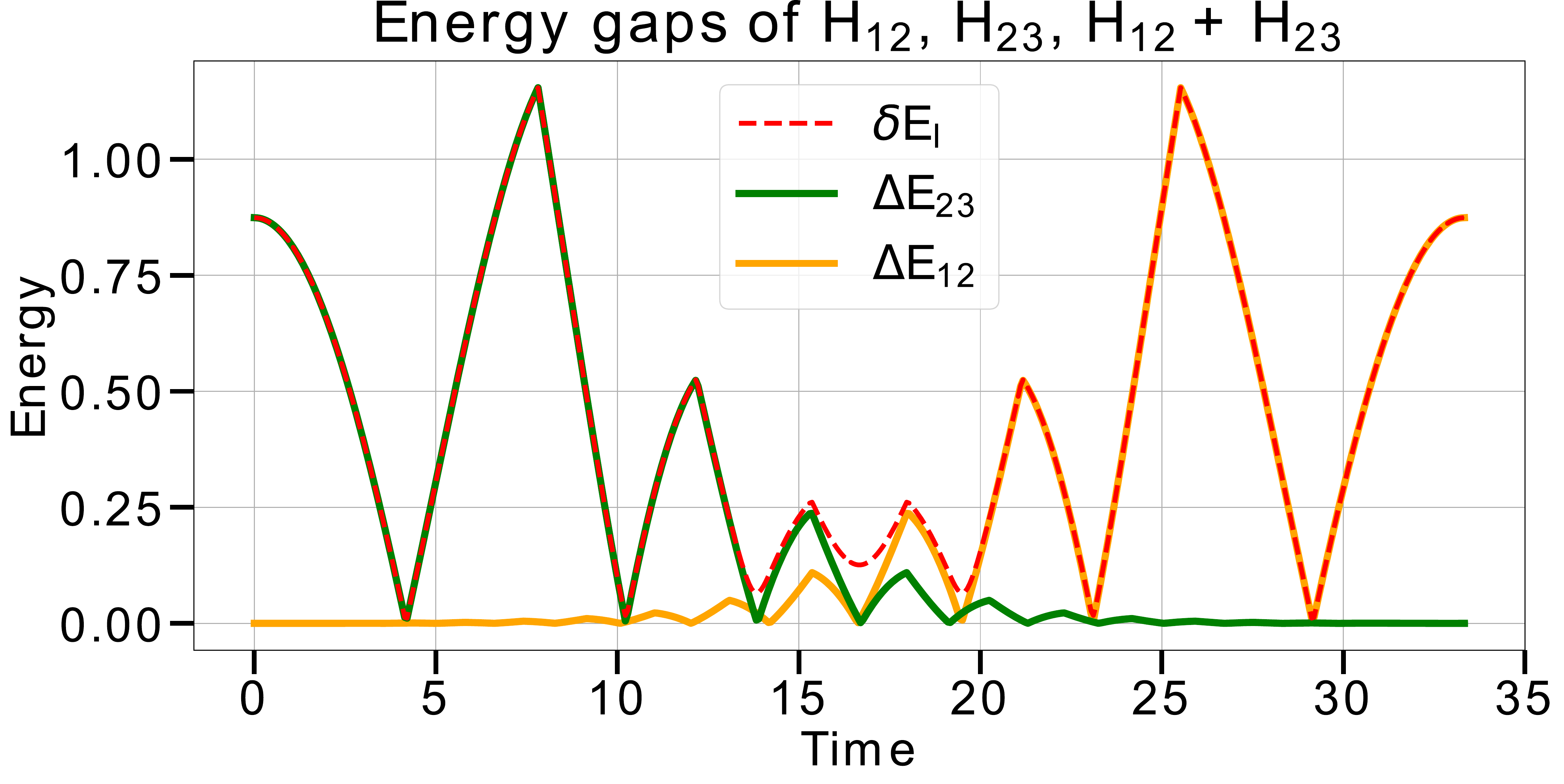}
\caption{\textit{Left panel:} Energy spectrum of the individual Hamiltonians $H_{12}$ and $H_{23}$, for $\mu =2$. Crossings occur in the spectra due to oscillations. \textit{Right panel:} Energy gap between the ground state and the first excited state of $H_{12}$, $H_{23}$ and $H_{12}+H_{23}$. In $H_{12}+H_{23}$, $H_{12}$ lifts the degeneracy of the crossings of $H_{23}$ and vice versa. The gap sizes are of the order $e^{-L_{ij}(\tau)}$. }
\label{fig:appendix3A}
\end{figure}

The origin of the avoided crossing in the simultaneous exchange protocol can be investigated by only considering the first stage Hamiltonian
\begin{equation}\label{eq:stage1Ham}
H_I(\tau) = t_{23}(\tau)\chi_2\chi_3^\dag+t_{12}(\tau)\chi_1\chi_2^\dag+ \hc = H_{23}(\tau)+H_{12}(\tau),\quad \tau \in [0,\tau_1]
\end{equation}
as well as the individual Hamiltonians $H_{12}$ and $H_{23}$ separately. In the left panel of Fig.~\ref{fig:appendix3A} we plot the individual spectrum of $H_{12}(\tau)$ and $H_{23}(\tau)$. The finite chemical potential induces oscillations and degeneracies which appear throughout the spectra. The amplitude of these oscillations are controlled by the exponential prefactor $e^{-L_{ij}}$. Consider now the gap between the ground state and the first excited state, which we plot in the right panel of Fig.~\ref{fig:appendix3A}. In $H_{23}$, due to the oscillations, the gap will vanish at certain points, before converging to zero due to the exponential suppression for increasing times. When the gaps of $H_{23}$ are exactly zero, which occurs during the period where the parafermionic pair $\chi_{2,3}$ is still coupled, the energy levels of $H_{12}$ are exponentially small. The opposite argument holds for the gaps of $H_{12}$.

In the right panel of Fig.~\ref{fig:appendix3A}, we show that the gap $\delta E_I(\tau)$ of the full Hamiltonian has the form $\delta E_I(\tau)  \geq  \text{max}\{\Delta E_{23}(\tau),\Delta E_{12}(\tau)\}$. This means that whenever the gap of $H_{12}(\tau)$ closes due to the oscillations, $H_{23}(\tau)$ will act as a perturbation and lift the degeneracy, and vice versa. In order to minimize the non-adiabatic transitions we can increase the total gap $\delta E_I(\tau)$ by keeping one of the gaps of the individual Hamiltonian as large as possible. This can be performed by keeping the corresponding parafermionic pair $\chi_{2,3}$ coupled as one approaches the other parafermion $\chi_1$ in order to couple $\chi_{1,2}$. The exchange becomes sequential rather than simultaneous. The same argument will also hold for the remainder of the braiding process.

An analytical expression for the gap with finite chemical potential can be found by mapping the first stage Hamiltonian~(\ref{eq:stage1Ham}) to a $\mathbb{Z}_4$ clock model using the Fradkin-Kadanoff transformation~(\ref{eq:fradkin}). This gives rise to a $16\times 16$ matrix which can be block diagonalized by rotating the Hamiltonian in the parity basis. The resulting four $4\times 4$ block Hamiltonians belong then to a given parity. The spectra for are the same for all parities, so we may choose the first of these resulting $4\times 4$ Hamiltonian matrix, which corresponds to the parity 0.
It reads:
\begin{align}
&\quad\tilde{H}_I(\tau)  =\nonumber\\
& \sqrt{2} |t_{23}(\tau)|\begin{pmatrix}
 -\cos [\varphi_{23}(\tau)]-\sin [\varphi_{23}(\tau)] & 0 & 0 & 0 \\
 0 &  \sin [\varphi_{23}(\tau)] - \cos [\varphi_{23}(\tau)] & 0 & 0 \\
 0 & 0 & \cos [\varphi_{23}(\tau)]+\sin [\varphi_{23}(\tau)] & 0 \\
 0 & 0 & 0 & \cos [\varphi_{23}(\tau)]-\sin [\varphi_{23}(\tau)]\nonumber \\
 \end{pmatrix}\\
& + |t_{12}(\tau)| \begin{pmatrix}
 0 & -e^{3 i \pi /4+i \varphi_{12}(\tau)} & 0 & e^{i \pi /4-i \varphi_{12}(\tau)} \\
 e^{i \pi /4-i \varphi_{12}(\tau)} & 0 & -e^{3 i \pi /4+i \varphi_{12}(\tau)} & 0 \\
 0 & e^{i \pi /4-i \varphi_{12}(\tau)} & 0 & -e^{3 i \pi /4+i \varphi_{12}(\tau)} \\
 -e^{3 i \pi /4+i \varphi_{12}(\tau)} & 0 & e^{i \pi /4-i \varphi_{12}(\tau)} & 0 \\
 \end{pmatrix},
\end{align}
where $\varphi_{ij}(\tau)=\arg[t_{ij}(\tau)]=\varphi_0+ \mu L_{ij}(\tau)$. For $\tau< \tau_1/2$, the parafermion pair $\chi_{2,3}$ is strongly coupled and thus $|t_{23}(\tau)|\gg |t_{12}(\tau)|$.
Avoided crossings will then appear in the spectra between the ground state and first excited state at time $\tau_c$ when
\begin{equation}\label{eq:avoidphase}
 \varphi_0+\mu L_{23}(\tau_c)  = \frac{\pi\lambda}{2},\quad \lambda\in\mathbb{Z}.
\end{equation}
By numerically computing the eigenvalues at these times we find that the gap between the ground state and the first excited state at $\tau_c$ reads:
\begin{equation}
\delta_\mu(\tau_c) \approx \sqrt{2}|t_{23}(\tau_c)|\left[\sqrt{1+ \sqrt{2} \frac{|t_{12}(\tau_c)|}{|t_{23}(\tau_c)|}}-\sqrt{1- \sqrt{2} \frac{|t_{12}(\tau_c)|}{|t_{23}(\tau_c)|} }\right] \approx 2e^{-L_{12}(\tau_c)},
\end{equation}
where we Taylor expanded the square roots to first order in $\sqrt{2} |t_{12}(\tau_c)|/|t_{23}(\tau_c)|$.
As expected, this gap goes to zero in the limit where $|t_{12}|\to 0$.

The smallest gap of the system $\delta(\mu)=\delta_\mu(\tau_{c})$ is obtained by solving Eq.~(\ref{eq:avoidphase}) for $\lambda=1$.
The case when $\tau>\tau_1/2$ is done by replacing $L_{12}\leftrightarrow L_{23}$, and the gap goes to zero when $|t_{23}|\to 0$ . As an example, the first gap in Fig.~\ref{fig:specFidB} appears at $\tau_{c}\approx 2.09$ with $\delta(\mu)\approx 9.5\times 10^{-4}$.

\end{document}